\documentclass[letterpaper,12pt]{article}

\usepackage{endnotes}
\usepackage{amssymb}
\usepackage{amsfonts}
\usepackage{amsmath} 
\usepackage{amsthm}
\usepackage{url}
\usepackage{graphicx,color}

\usepackage{soul}
\usepackage{lscape}

\usepackage{comment}
\usepackage{chicago}
\usepackage{array}

\usepackage{url} 

\usepackage{setspace}
\newcommand{\hh}{\hat{h}}
\newcommand{\hT}{\hat{T}}

\def\R{\mathbb{R}}
\def\endproof{\hfill\diamondsuit}

\def\sF{{\mathcal F}}

\def\sA{{\mathcal A}}

\def\E{\mathbb{E}}
\def\V{\mathbb{V}}
\def\sF{\mathcal{F}}
\def\P{\mathbb{P}}

\def\N{\mathbb N}

\numberwithin{equation}{section}
\theoremstyle{plain}                
\newtheorem{theorem}{Theorem}[section]
\newtheorem{lemma}[theorem]{Lemma}
\newtheorem{proposition}[theorem]{Proposition}

\theoremstyle{definition}           
\newtheorem{definition}[theorem]{Definition}

\theoremstyle{remark}               
\newtheorem{remark}{Remark}[section]

\usepackage[margin=1.25in]{geometry}

\usepackage{ulem} 

\begin{document}
\pagestyle{empty}

\begin{center}
\large{\bf Resolving asset-pricing puzzles using price-impact}$^\ast$
\makeatletter{\renewcommand*{\@makefnmark}{}\footnotetext{\hspace{-.35in} $^\ast${The authors have benefited from helpful comments from Yashar Barardehi, Suleyman Basak, Ren\'e Carmona, Joel Hasbrouck, Burton Hollifield, Ioannis Karatzas, Lars Kuehn, Bryan Routledge, Mete Soner,  and seminar participants at Tepper (Carnegie Mellon), ORFE (Princeton), and Intech. The third author has been supported by the National Science Foundation under Grant No. DMS 1812679 (2018 - 2021). Any opinions, findings, and conclusions or recommendations expressed in this material are those of the author(s) and do not necessarily reflect the views of the National Science Foundation (NSF). The corresponding author is Jin Hyuk Choi. Xiao Chen has email: xc206@math.rutgers.edu, Jin Hyuk Choi has email: jchoi@unist.ac.kr,  Kasper Larsen has email: KL756@math.rutgers.edu, and Duane J. Seppi has email: ds64@andrew.cmu.edu. }\makeatother}}
\end{center}

\begin{center}

{ \bf Xiao Chen}\\
Rutgers University

\ \\ 

{ \bf Jin Hyuk Choi}\\ 
Ulsan National Institute of Science and Technology (UNIST) 

\ \\

{ \bf Kasper Larsen}\\
Rutgers University

\ \\ 

{ \bf Duane J. Seppi}\\
Carnegie Mellon University

\end{center}
\begin{center}

{\normalsize \today }
\ \\
\end{center}

\begin{verse}
{\sc Abstract}: We solve in closed-form an equilibrium model in which a finite number of exponential investors continuously consume and trade with price-impact. Compared to the analogous Pareto-efficient equilibrium model, price-impact has an amplification effect on risk-sharing distortions that helps resolve the interest rate puzzle and the stock-price volatility puzzle and, to a lesser extent, affects the equity premium puzzle.

\end{verse}

\begin{verse}
{\sc Keywords}: Asset pricing, price-impact, Nash equilibrium, Radner equilibrium, risk-free rate puzzle, equity premium puzzle, volatility puzzle.

\ \\
\end{verse}
\newpage

\pagestyle{plain}
\addtocounter{page}{-1}

\section{Introduction}

Understanding the effects of market frictions on pricing and trading is a long-standing topic of interest in financial economics.  The market microstructure literature focuses on informational frictions and liquidity-provision frictions (e.g., Kyle 1985, Stoll 1978, Grossman and Miller 1998). In contrast, the consumption-based asset pricing literature studies how various frictions affect risk-sharing across investors and, thus, affect interest rates, stock-price volatility, and the market price-of-risk.\footnote{Previous research shows that model incompleteness and consequent non-efficient risk-sharing equilibria can arise from several channels including: (i) Unspanned labor income as in the continuous-time Radner models in Christensen, Larsen, and Munk (2012),  {\v Z}itkovi\'c (2012), Christensen and Larsen (2014), Choi and Larsen (2015), Kardaras, Xing, and  {\v Z}itkovi\'c  (2015), Larsen and Sae-Sue (2016), and \ and {\v Z}itkovi\'c (2020). (ii) Limited stock-market participation and trading constraints as in the continuous-time Radner models in Basak and Cuoco (1998) and Hugonnier (2012). (iii) Transaction costs and quadratic penalties as in the Radner models in Heaton and Lucas (1992, 1996), Vayanos and Vila (1999), G\^arleanu and Pedersen (2016), Bouchard, Fukasawa, Herdegen, and Muhle-Karbe (2018), and Weston (2018). (iv) Trading targets as in the continuous-time Nash models in Brunnermeier and Petersen (2005), Sannikov and Skrzypacz (2016), and Choi, Larsen, and Seppi (2020). (v) Price-impact as in the discrete-time Nash model in Vayanos (1999) and the continuous-time Nash models in Basak (1996, 1997).
} This paper investigates the asset-pricing effects of strategic investor behavior with price-impact frictions on continuous-time stock-price dynamics and interest rates.    

Much of our modeling approach is standard. A finite number of risk-averse investors with time-separable utility receive individual income over time and trade a stock that pays exogenous continuous dividends and a money market account. Consumption and trading decisions occur in continuous time over a finite time horizon. Investors trade due to initial stock-holding endowment imbalances.  The key innovation in our model is that investors are strategic with respect to the perceived price-impact of their asset holdings and trades. Our main theorem provides the Nash equilibrium stock-price process and equilibrium interest rate with price-impact via solutions to a system of ODEs. 

Our main application shows that price-impact in our Nash equilibrium model has material effects on the equilibrium interest rate and stock-price process relative to both the analogous competitive price-taking Radner equilibrium (with unspanned income shocks and no price-impact) and the analogous Pareto-efficient equilibrium (with spanned income shocks and without price-impact). 
More specifically, taking the Pareto-efficient equilibrium model as a baseline, price-impact in our Nash equilibrium model magnifies risk-sharing distortions and, as a result, can simultaneously lower the interest rate, increase stock-price volatility, and, to a lesser extent, increase the equity premium. Therefore, price-impact can simultaneously help resolve the risk-free rate puzzle of Weil (1989), the volatility puzzle of LeRoy and Porter (1981) and Shiller (1981), and marginally affects with the equity premium puzzle of Mehra and Prescott (1985). To the best of our knowledge, it is a new insight that price-impact can matter for these asset-pricing puzzles.


A variety of other approaches have been proposed to resolve the three asset-pricing puzzles: 
(i) Constantinides and Duffie (1996) and variations including Storesletten, Telmer, and Yaron (2007, 2008) and Krueger and Lustig (2010) use permanent idiosyncratic income shocks to resolve the three asset-pricing puzzles. 
However, Cochrane (2005, p.478-9) argues that high levels of risk aversion are still needed to explain the equity premium puzzle  in Constantinides and Duffie (1996). Furthermore, Cochrane (2008, p.310) argues that the continuous-time limiting model of Constantinides and Duffie (1996) requires jumps to explain the puzzles.\footnote{A closed-form competitive Radner equilibrium model with exponential utility investors and dividend and income processes governed by continuous-time L\'evy jump processes which can simultaneously explain the three puzzles is given in Larsen and Sae-Sue (2016).}  In contrast to approach (i), our price-impact equilibrium model has modest levels of risk aversion and no jumps. In particular, we use correlated arithmetic Brownian motions to generate exogenous stock dividends and  strategic investor idiosyncratic income shocks.\footnote{Additionally, Judd (1985), Feldman and Gilles (1985), and Uhlig (1996) present both mathematical and interpretation  issues related to models with a continuum of investors --- such as Constantinides and Duffie (1996) --- because these models rely on average clearing conditions. In contrast, our equilibrium model's idiosyncratic income shocks persist at the aggregate level. }
(ii) In a representative agent framework,  Constantinides (1990) uses  an internal habit process and  Campbell and Cochrane (1999) use an external habit process to explain the puzzles. 
(iii) Bansal and Yaron (2004) combine long-run consumption risk and an Epstein-Zin  representative agent to explain the puzzles. 
(iv)  Barro (2006) and the extension to an Epstein-Zin representative agent  in Wachter (2013) use rare disasters based on jump processes  to resolve the puzzles. 
In contrast to approaches (ii)-(iv), our investors' utilities are time-additive separable exponential utility functions over continuous-time consumption rate processes. Furthermore, the models in  approaches (ii)-(iv) are based on representative-agent frameworks in which the underlying model is effectively complete. However, our model incorporates unspanned income shocks and price-impact. Of the models in (ii)-(iv), our model is closest to the external habit model in Campbell and Cochrane (1999). Indeed, by switching off our model's idiosyncratic income shocks, the resulting common income shocks can be interpreted as an external habit.

A non-standard feature in our analysis is that it is non-stationary in that the asset pricing effects of price-impact dissipate over time. In our model, investors start with endowed initial stock positions that are Pareto inefficient.  However, due to price-impact, investors do not trade immediately to efficient risk-sharing; rather they trade gradually to optimize with respect to a trade-off between the benefits of improved risk-sharing and price-impact costs of faster trading.  Over time, their gradual trading has a cumulative effect that improves risk-sharing.  Thus, our analysis shows that price-impact can have a quantitatively material short-term amplification effect on asset pricing by prolonging risk-sharing distortions.  In our model, risk sharing distortions arise as a one-time occurrence via inequalities in initial endowed stock positions. In richer economic settings, however, risk-sharing distortions could arise on a reoccurring basis from stochastic habits, income shocks, heterogeneous beliefs, and asymmetric information. In such a reoccurring-shock environment, the asset pricing amplification effect due to price-impact could be part of asset pricing in a stationary equilibrium. Moreover, from a calibration perspective, fundamental risk-bearing shocks can be quantitatively smaller (i.e., more realistic) and still have material asset pricing effects because they would be magnified by the price-impact amplification effect.

Basak (1996, 1997), Vayanos (1999, 2001), and Pritsker (2009) develop equilibrium models with price-impact.
The main differences between our model and Basak (1996, 1997) are: First, unspanned income shocks make our model incomplete. Second, we allow for multiple traders with price-impact. Third, our price-impact equilibrium model is time-consistent. Our analysis extends or differs from Vayanos (1999, 2001) and Pritsker (2009) in three ways: First, we solve for an endogenous deterministic interest rate. Among other things, this allows us to investigate the interest rate puzzle in Weil (1989). In particular, we find that price-impact has a quantitatively  larger effect on endogenous interest rates than on the equity Sharpe Ratio. Second, our investors start with non-Pareto efficient initial stock endowments, but then subsequently receive stochastic income shocks rather than stock-holding shocks. Third, and more technically, our model is in continuous time, which makes the analysis mathematically tractable.\footnote{While Vayanos (2001) allows for exogenous noise traders, all our investors are utility maximizers.}

Optimal portfolio and consumption choice in models with price-impact and in models with transaction costs often produce similar implications for optimal investor behavior, but there is a key difference for asset pricing. This is because all markets must clear in equilibrium. For non-monetary models (such as ours), transaction costs complicate the clearing condition for the real good market because transaction costs paid by one investor must be consumed by others. The price-impact mechanism we use is standard (see, e.g., Vayanos 1999) and does not affect any clearing conditions.  In particular, price-impact in realized prices is a form of price pressure in the prices paid and received by buyers and sellers, rather than separate auxiliary cash flows as, for example, transaction costs. In addition, perceived price-impact is an investor perception whereas clearing conditions must hold for realized investor behavior. 

Lastly, our analysis is related to a long-standing question in financial economics about whether liquidity is priced (see, e.g., surveys in Easley and O'Hara (2003) and Amihud, Mendelson, and Pedersen (2006)).  One literature holds that liquidity is priced because investors require compensation for holding securities that expose them to higher transaction costs. For example, Amihud and Mendelson (1986) provide a theoretical analysis of this effect.  Acharya and Pedersen (2005) also show that systemic uncertainty in stochastic trading costs (seen as a type of random negative dividends) can be a priced risk factor. However, another literature argues that the quantitative asset pricing impact of liquidity is small by showing in various economic settings that investors can reduce their trading with only small utility costs. This counter-argument was first presented in Constantinides (1986).  In contrast, our model is not about bid-ask spreads and transactional forms of illiquidity, but rather about the price-impact of investor asset-demand imbalances on market-clearing prices. In particular, we show, in an analytically tractable version of a standard general equilibrium asset pricing framework, how persistent distortions in risk-sharing due to how investors curtail their trading in response to price-impact  has asset pricing effects.

The paper is organized as follows: Section \ref{Sec1} sets up the individual optimization problems including the perceived price-impact functions. Section \ref{Sec:2} contains our main theoretical result, which provides our price-impact Nash equilibrium in closed-form. Section  \ref{sec:app} shows in numerical examples how price-impact can simultaneously affect the risk-free interest rate puzzle, the equity premium puzzle, and the volatility puzzle. Appendix \ref{app:pf} contains proofs,  Appendix \ref{ssecC} outlines  the analogous competitive Radner model, and Appendix \ref{ssecR} uses the consumption-based CAPM to derive the analogous Pareto-efficient equilibrium in closed-form. Appendix \ref{app:cal} discusses price-impact calibration.

\section{Setup}\label{Sec1}

We consider a real economy model with a single perishable consumption good, which we take as the model's num\'eraire. Trading and consumption take place continuously for $t\in[0,T]$ for a finite time-horizon $T\in(0,\infty)$. The model has two traded securities: A money market account and a stock. The money market account is in zero net supply, and the stock supply is a constant $L\in\N$. The stock pays exogenous random dividends given by a rate process $D=(D_t)_{t\in[0,T]}$ per share.  The investors receive income given by exogenous random rate processes $Y_i =(Y_{i,t})_{t\in[0,T)}$ for $i\in\{1,...,I\}$ for  $I\in\N$. In Theorem \ref{thm:Nash} below, we determine endogenously the interest rate $r =\big(r(t)\big)_{t\in[0,T]}$ (a deterministic time-varying function) and the stock-price process $\hat{S}=(\hat{S}_t)_{t\in[0,T]}$ in a Nash equilibrium with price-impact.

\subsection{Exogenous model inputs} 
Let $(B_t, W_{1,t},...,W_{I,t})_{t\in[0,T]}$ be independent one-dimensional Brownian motions starting at zero with zero drifts and unit volatilities. The augmented standard Brownian filtration is denoted by
\begin{align}\label{sFi}
\sF_{t}:=\sigma(B_s,W_{1,s},...,W_{I,s})_{s\in[0,t]},\quad t\in[0,T].
\end{align}

An exogenous stock dividend rate process $D_t$ has dynamics
\begin{align}\label{dD}
dD_t := \mu_D dt + \sigma_D dB_t, \quad D_0 \in \R,
\end{align}
driven by the Brownian motion $B_t$, with a given initial value $D_0$, a constant drift $\mu_D$, and a constant volatility coefficient $\sigma_D \geq 0$.
The dividend rate process plays two roles: First, it generates a running flow of instantaneous dividends where the associated cumulative dividend over $[0,t]$ is $\int_0^t D_s ds$ for $t \in [0, 1]$. 
Second, the stock pays a final dividend $D_T$ at the terminal date $T$ that pins down the terminal stock price: 
\begin{align}\label{pp10}
\lim_{t\uparrow T} S_t =D_T,\quad \P\text{-a.s.}
\end{align}
The terminal condition \eqref{pp10}  requires the stock-price process $S = (S_t)_{t\in[0,T]}$ to be left-continuous at time $t = T$. We refer to Ohasi (1991, 1992) for a discussion of \eqref{pp10}. A boundary condition like \eqref{pp10} is needed since our model, for mathematical tractability, has a finite time horizon.  However, by making $T$ large, the terminal liquidating dividend $D_T$ is small relative to total dividends $\int_0^T D_s ds + D_T$. In the next section, we require that \eqref{pp10} holds for both investor $i$'s perceived stock-price process $S = S_i$ (to be defined in Section \ref{sec2_3}) and for the equilibrium stock-price process $S = \hat S$ (to be proven to exist in Section \ref{Sec:2}).

The Brownian motion $W_{i,t}$ generates idiosyncratic income shocks in the income rate process $Y_{i,t}$ for trader $i\in\{1,...,I\}$. We model $Y_{i,t}$ as in Christensen, Larsen, and Munk (2012) and define
\begin{align}\label{dYi}
dY_{i,t} := \mu_Y dt + \sigma_Y\big(\rho  dB_t+\sqrt{1-\rho^2}dW_{i,t}\big), \quad Y_{i,0} \in \R.
\end{align}
Investor $i$'s income consists of a flow of income over $[0,T]$ resulting in cumulative income given by $\int_0^T Y_{i,s}ds$ and then a lump-sum income payment $Y_{i,T}$ at the end. The terminal payment $Y_{i,T}$ is a reduced-form for the value of a flow of income after the terminal date $T$.  Similar to the boundary condition for dividends, the terminal lump-sum income $Y_T$ can be made small relative to total income $\int_0^T Y_s ds + Y_T$ by making $T$ large. In the income rate dynamics \eqref{dYi}, the given  initial value is $Y_{i,0}$, the constant drift is $\mu_Y$, the constant volatility coefficient is $\sigma_Y\ge0$, and $\rho \in [-1,1]$ is a correlation parameter controlling the relative magnitudes of investor-specific (i.e., idiosyncratic) income shocks and income shocks correlated with the dividend process in \eqref{dD}. For example, $\rho:=0$ makes all income shocks independent of dividend shocks.  When $\rho^2<1$ in \eqref{dYi}, no single stock-price process can span all risk because any model with multiple Brownian motions and only one stock is necessarily incomplete by the Second Fundamental Theorem of Asset Pricing. However, when $\rho^2=1$, all randomness in the model is due to the Brownian motion $B_t$, and model completeness is possible. While the assumption of homogenous income coefficients is common in many Nash equilibrium models, Section  \ref{sec:extension} below considers an extension with heterogenous investor income coefficients.

We model the asset-holding decisions of a group of $j\in \{1,...,I\}$, $I\in\N$, strategic traders.  We normalize the strategic traders' endowed money market balances to zero. Traders begin with exogenous initial individual stock endowments equal to constants $\theta_{j,0}\in\R$ for $j\in \{1,...,I\}$. Their stock-holding processes over time are 
\begin{align}\label{ad1}
\begin{split}
\theta_{j,t} &:= \theta_{j,0}+\int_0^t\theta'_{j,u}du,\quad t\in[0,T].
\end{split}
\end{align}
This restriction forces traders to use only holding processes given by continuous order-rate processes $\theta'_{j,t}$. This rate-process  restriction has been used in various equilibrium models including Back, Cao, and Willard (2000), Brunnermeier and Pedersen (2005), G\^arleanu and Pedersen (2016), and Bouchard, Fukasawa, Herdegen, and Muhle-Karbe (2018).  In Section  \ref{sec:extension} below we show how to incorporate discrete orders (i.e., block orders) into the model. 

At time $t\in[0,T]$, trader $i$ chooses an order-rate process $\theta'_{i,t}$ and a consumption rate process $c_{i,t}$. In aggregate, these processes clear the stock and real-good consumption  markets in the sense that
\begin{align}\label{clearing100}
L= \sum_{i=1}^{I} \theta_{i,t},\quad LD_t + \sum_{i=1}^IY_{i,t}=\sum_{i=1}^Ic_{i,t},\quad t\in[0,T],
 \end{align} 
where $L$ is the constant stock supply. Walras' law ensures that clearing in the stock and real-good consumption markets lead to clearing in the zero-supply money market. The terminal stock price \eqref{pp10} ensures clearing in the real good consumption market at the terminal time $T$.

Our model is constructed to investigate how price-impact affects risk-sharing and, thus, asset pricing.   Two specific types of risk-sharing distortions are present in the model:  The first is potential deviations of investors' initial endowments $\theta_{i,0}$ from equal holdings $\frac{L}{I}$. A second distortion is unspanned stochastic investor income.  Section \ref{Sec:2} investigates both distortions.


\subsection{Individual utility-maximization problems} 

With price-impact in our model, traders perceive that their holdings $\theta_{i,t}$ and order rates $\theta_{i,t}'$ affect the prices at which they trade and their resulting wealth dynamics. In particular, price-impact here is due to the impact of investor holdings on the market-clearing aggregate risk-bearing capacity of the market, and a microstructure impact of investor trading. Trader $i$'s perceived wealth process is defined by
\begin{align}\label{X}
X_{i,t} :=\theta_{i,t} S_{i,t} + M_{i,t},\quad t\in[0,T],\quad i\in\{1,...,I\},
\end{align}
where $\theta_{i,t}$ denotes her stock holdings, $S_{i,t}$ is her perceived stock-price process, and $M_{i,t}$ is her money-market balance (all these processes are to be determined in equilibrium endogenously).  In a Nash equilibrium model, the perceived stock-price processes $S_{i,t}$ in \eqref{X} can differ off-equilibrium across traders given their different hypothetical holdings $\theta_{i,t}$ and trades $\theta'_{i,t}$ but the equilibrium stock-price process $\hat{S}_t$ is identical for all traders. On the other hand, we assume all traders perceive the same deterministically time-varying interest rate $r(t)$, $t\in[0,T]$ (to be determined  endogenously).

Recall that we have normalized each strategic trader's initial money market account balance to zero whereas the initial endowed stock holdings are exogenously given by $\theta_{i,0}\in \R$. The self-financing condition produces trader $i$'s perceived wealth dynamics 
\begin{align}\label{dX}
d X_{i,t} =  r(t)M_{i,t}dt +\theta_{i,t} (dS_{i,t}+D_tdt)+(Y_{i,t} -c_{i,t})dt,\quad X_{i,0}= \theta_{i,0}S_{i,0}.
\end{align}

As usual in continuous-time stochastic control problems, the traders' controls must satisfy various regularity conditions. 
\begin{definition}[Admissibility]\label{ad} An order-rate process $\theta_i'= (\theta'_{i,t})_{t\in[0,1]}$ and a consumption-rate process  $c_i= (c_{i,t})_{t\in[0,1]}$ are admissible, and we write $(\theta'_i,c_i) \in \sA$ if:
\begin{enumerate}
\item[(i)] The processes $(\theta_i',c_i)$ have continuous paths and are progressively measurable with respect to the filtration $\sF_{t}$ in \eqref{sFi}.

\item[(ii)] The stock-holding process $\theta_{i,t}$ defined by \eqref{ad1} is uniformly bounded. 
\item [(iii)] The wealth process dynamics \eqref{dX} as well as the corresponding money market account balance process $M_{i,t} := X_{i,t}-S_{i,t}\theta_{i,t}$ are well-defined and
\begin{align}\label{mmg}
\sup_{t\in[0,T]}\E[e^{\zeta M_{i,t}}] <\infty\quad \text{for all}\quad \zeta\in \R.
\end{align}
\item [(iv)] The perceived stock-price process $S_{i,t}$ satisfies the terminal condition \eqref{pp10}.
\end{enumerate}
$\endproof$
\end{definition}

Each trader $i$ seeks to solve\footnote{The negative sign in the exponential utility is removed for simplicity, which leads to the minimization problem in \eqref{optproblem}.
}
\begin{align}\label{optproblem}
\inf_{(\theta'_{i},c_{i})\in \sA} \E\left[\int_0^T e^{-a c_{i,t} -\delta t}dt + e^{-a (X_{i,T}+Y_{i,T})-\delta T}\right],\quad i=1,...,I,
\end{align}
given the perceived stock-price process $S_{i,t}$ in her wealth dynamics 
\eqref{dX}. In \eqref{optproblem}, the term $\int_0^T e^{-a c_{i,t} -\delta t}dt$ denotes utility from the consumption flow rates and the term $e^{-a (X_{i,T}+Y_{i,T})-\delta T}$ is a bequest value function for terminal wealth. Like the terminal dividend $D_T$ and the lump-sum terminal income $Y_T$, the bequest utility function proxies the continuation utility past the terminal time in our model.  
 For tractability, the common absolute risk-aversion coefficient $a>0$ is the same for both the consumption flow utility and the bequest value function. The common time preference parameter is $\delta\ge0$. The assumption of homogenous exponential utilities is common in many Nash equilibrium models, see, e.g., Vayanos (1999). In Section \ref{sec:extension} below we allow for heterogenous exponential utilities across investors.

The next subsection derives stock-price dynamics  perceived by trader $i$  when solving \eqref{optproblem} as part of our Nash equilibrium with price-impact.  These perceived price dynamics differ from those in the competitive Radner and Pareto-efficient equilibria where all traders perceive the same stock-price and act as price-takers. We describe the analogous competitive Radner equilibrium in Appendix \ref{ssecC} and the analogous competitive Pareto-efficient equilibrium  in Appendix \ref{ssecR}. As is shown below, neither our Nash model with price-impact nor the analogous competitive Radner model is Pareto efficient. In addition, when the idiosyncratic income shocks are turned off, the Radner equilibrium reduces to the Pareto-efficient equilibrium whereas our Nash model remains Pareto inefficient due to price-impact.

\subsection{Price-impact for the stock market}\label{sec2_3}
The perceived stock-price process $S_{i,t}$ for trader $i$ depends on market-clearing given how the other traders $j\in \{1,...,I\}\setminus \{i\}$ respond to trader $i$'s hypothetical choices  of $\theta'_{i,t}$. Thus, for a Nash equilibrium, we must model how traders $j$, $j\neq i$,  respond to an arbitrary control $\theta'_{i,t}$ used by trader $i$. 

Several different price-impact models are available in the literature: Kyle (1985) and Back (1992) use continuous-time price-impact functions in which price changes $dS_{i,t}$ are affine in orders $d\theta_{i,t}$. Cvitani\' c and Cuoco (1998) take the drift process in $dS_{i,t}$ to be a function of $\theta_{i,t}$. The affine price-impact function \eqref{optimal101} we derive 
below can be found in the single-trader optimal order-execution models in  Almgren (2003) and Schied and Sch\"oneborn (2009). Our Nash equilibrium model with price-impact 
can be seen as a continuous-time version of the discrete-time Nash equilibrium model in Vayanos (1999) where $S_{i,t_n}$ is affine in discrete orders $\Delta \theta_{i,t_n}$.

For a fixed trader with index $i\in\{1,...,I\}$, we conjecture that the perceived responses used by other traders $j$, $j\neq i$, to hypothetical holdings $\theta_{i,t}$ and trades $\theta'_{i,t}$ by investor $i$ are given by
\begin{align}\label{optimal100}
\begin{split}
\theta'_{j,t} &:=  A_0(t)\big(F(t)D_{t}-S_{i,t}\big)+A_1(t)\theta_{j,t}+ A_2(t)\theta_{i,t}+ A_3(t)\theta_{i,t}',\quad j\neq i,
\end{split}
\end{align}
for deterministic functions of time $A_0(t),...,A_3(t)$. The intuition behind \eqref{optimal100} is that investors $j\neq i$ are perceived by investor $i$ to have base levels for their orders $\theta'_{j,t}$ that they then adjust given the controlled price level $S_{i,t}$ (which is affected by trader $i$'s holdings  $\theta_{i,t}$  and orders $\theta_{i,t}'$) 
relative to an adjusted dividend level $F(t)D_t$ where $F(t)$ is the annuity\footnote{For future reference, note that \eqref{ann} is equivalent to $F(T)=1$ and $F'(t)=r(t)F(t)-1$.}
\begin{align}\label{ann}
F(t) := \int_t^T e^{-\int_t^sr(u)du} ds + e^{-\int_t^Tr(u)du},\quad t\in[0,T].
\end{align}
The response specification in \eqref{optimal100} also allows the perceived responses of investors $j \neq i$ to depend directly on investor $i$'s hypothetical holdings $\theta_{i,t}$ and orders $\theta_{i,t}'$. Thus, $S_{i,t}$ is not assumed to be a sufficient statistic for the effects of $\theta_{i,t}$ and $\theta_{i,t}'$ on $\theta_{j,t}'$.  At the end of this subsection, we show that \eqref{optimal100} can be rewritten as trader $j$ deviating from $j$'s equilibrium behavior in response to trader $i$'s off-equilibrium behavior.

The perceived investor-response functions $A_0(t),..., A_3(t)$ in \eqref{optimal100} are not simply assumed. Rather, these functions are endogenously determined in equilibrium in  Theorem \ref{thm:Nash}  below given market-clearing, certain belief-consistency conditions (described in Definition \ref{Nash}  below), and given a microstructure parameter that implicitly determines the temporary (transitory) price-impact of trading.  
 
The stock-price process $S_{i,t}$ trader $i$ perceives in her optimization problem \eqref{optproblem} is found using the stock-market clearing conditions \eqref{clearing100} given the perceived responses in \eqref{optimal100}:
\begin{align}\label{optimal88}
\begin{split}
0 &= \theta_{i,t}'+\sum_{j\neq i} \theta_{j,t}'\\
&=\theta'_{i,t}+ (I-1)A_0(t)\big(F(t)D_{t}-S_{i,t}\big)\\  &\quad+A_1(t)(L-\theta_{i,t})+(I-1)\big(A_2(t)\theta_{i,t}+ A_3(t)\theta_{i,t}'\big).
\end{split}
\end{align}
Provided that $A_0(t)\neq 0$ for all $t\in[0,T]$, we can solve \eqref{optimal88} for trader $i$'s perceived market-clearing stock-price process:
\begin{align}\label{optimal101}
\begin{split}
S_{i,t}&= D_t F(t)+ \tfrac{A_1(t) L}{A_0(t) (I-1)} + \tfrac{A_2(t)(I-1)-A_1(t)}{A_0(t) (I-1)}\theta_{i,t}+ \tfrac{A_3(t) (I-1)+1}{A_0(t) (I-1)}\theta_{i,t}'.
\end{split}
\end{align}
Trader $i$'s stock holdings $\theta_{i,t}$ and orders $\theta'_{i,t}$ affect the perceived stock-price process \eqref{optimal101} as follows. Similar to  Almgren (2003), the sum $F(t)D_t+ \frac{A_1(t)L}{A_0(t)(I-1)}$  in \eqref{optimal101} is called the \emph{fundamental stock-price process}.  The coefficient $\frac{A_2(t) (I-1)-A_1(t)}{A_0(t)(I-1)}$ on holdings $\theta_{i,t}$ in \eqref{optimal101} is the \emph{permanent} price-impact (positive in equilibrium)  because the price-impact effect of investor $i$'s  past trading persists even after trading stops   (when $\theta_{i,t}'=0$ and $\theta_{i,t}\neq0$). The coefficient $\frac{A_3(t) (I-1)+1}{A_0(t)(I-1)}$ on the order rate $\theta'_{i,t}$ in \eqref{optimal101} is the \emph{temporary} price-impact (positive in equilibrium)  because this component of the price-impact effect disappears when investor $i$ stops trading (i.e., when $\theta'_{i,t}=0$). Theorem \ref{thm:Nash} below provides $A_0(t),...,A_3(t)$ via solutions to a coupled system of ODEs. 

To see that \eqref{optproblem} is a quadratic minimization problem for the perceived stock-price process \eqref{optimal101}, we use the money-market account balance process $M_{i,t}$ from \eqref{X} defined by
\begin{align}\label{M}
M_{i,t}:= X_{i,t} -\theta_{i,t}S_{i,t},\quad i\in\{1,...,I\},
\end{align}
as a state-process. The wealth dynamics \eqref{dX} produce the following dynamics of the money-market account balance process
\begin{align}\label{dM}
\begin{split}
dM_{i,t}&= dX_{i,t} -d(\theta_{i,t} S_{i,t})\\
&= r(t)M_{i,t}dt + \theta_{i,t} (dS_{i,t}+D_tdt)+(Y_{i,t}- c_{i,t})dt- \theta'_{i,t} S_{i,t} dt -\theta_{i,t} dS_{i,t}\\
&= \Big(r(t)M_{i,t} +\theta_{i,t}D_t -S_{i,t}\theta_{i,t}' +Y_{i,t}- c_{i,t}\Big)dt.
\end{split}
\end{align}
The second equality in \eqref{dM} uses the quadratic variation property $\langle \theta_i,S_i\rangle_t =0$, which holds because $\theta_{i,t}$ satisfies the order-rate condition \eqref{ad1}. As shown in the proof in Appendix \ref{app:pf}, the affinity in the price-impact function \eqref{optimal101} and the last line in \eqref{dM} make the individual optimization problems \eqref{optproblem} tractable. 

Trader $i$'s control $\theta'_{i,t}$ appears implicitly in trader $j$'s response \eqref{optimal100} through the stock-price process $S_i = (S_{i,t})_{t\in[0,1]}$ and directly via $\theta_{i,t}$ and $\theta'_{i,t}$. Substituting \eqref{optimal101} into \eqref{optimal100},  the resulting response functions  for $j\neq i$ give trader $j$'s response directly in terms of trader $i$'s orders $\theta_{i,t}'$ and associated holdings $\theta_{i,t}$, where trader $j$'s response is affine in those quantities:
\begin{align}\label{response2}
\begin{split}
\theta'_{j,t}=A_1(t)\theta_{j,t}+\frac{A_1(t)}{1-I}(L- \theta_{i,t})+\frac{1}{1-I}\theta_{i,t}'.
\end{split}
\end{align}
 Furthermore, the equilibrium holdings  $(\hat{\theta}_{i,t},\hat{\theta}_{j,t})$ and order-rate processes $(\hat{\theta}'_{i,t},\hat{\theta}'_{j,t})$ in  Theorem \ref{thm:Nash} are consistent with \eqref{optimal100} in the sense
\begin{align}\label{optimal100b}
\hat{\theta}'_{j,t} &=  A_0(t)\big(F(t)D_{t}-\hat{S}_t\big)+A_1(t)\hat{\theta}_{j,t}+ A_2(t)\hat{\theta}_{i,t}+ A_3(t)\hat{\theta}_{i,t}',\quad j\neq i,
\end{align}
given the equilibrium stock-price process  $\hat{S}_t$. This allows us re-write  \eqref{response2} as
\begin{align}\label{response2222}
\begin{split}
\theta'_{j,t} &=  \hat{\theta}'_{j,t} -A_1(t)(\hat{\theta}_{j,t}-\theta_{j,t})+\frac{1}{I-1}(\hat{\theta}'_{i,t}-\theta_{i,t}')-\frac{A_1(t)}{I-1}(\hat{\theta}_{i,t}-\theta_{i,t}).
\end{split}
\end{align}
Thus, the responses in  \eqref{response2222} describe deviations of $\theta'_{j,t}$ from equilibrium behavior $\hat{\theta}'_{j,t}$ for trader $j$, $j\neq i$, in response to trader $i$'s off-equilibrium deviations of $\theta'_{i,t}$ from $\hat{\theta}'_{i,t}$. Note here that the equilibrium holdings $(\hat{\theta}_{i,t},\hat{\theta}_{j,t})$ and order-rate processes $(\hat{\theta}'_{i,t},\hat{\theta}'_{j,t})$, $j\neq i$, in \eqref{response2222} do not depend on trader $i$'s arbitrary orders $\theta_{i,t}'$ and holdings $\theta_{i,t}$. 

\subsection{Modeling approach}

This section briefly describes modeling differences between our analysis and other asset pricing models and explains the motivation and reasons for these differences. 

With pricing-taking exponential investors, the initial endowed stock-holding distribution across investors is irrelevant, as is well-known, in asset pricing models. However, our exponential investors are strategic in that they perceive their holdings and trades to have price-impact, which explains why our equilibrium model exhibits stock endowment effects. However, these endowment effects are due to a risk-bearing mechanism rather than a wealth effect. When investors are endowed with non-Pareto efficient initial stock endowments in terms of risk-sharing, it is suboptimal for investors to immediately trade to their Radner allocations due to their perceived costs given their perceived price-impact (when $\rho^2=1$ in \eqref{dYi}, the idiosyncratic income shocks disappear and the Radner equilibrium becomes Pareto efficient). The deviation of risk-sharing in the model relative to the Radner equilibrium, in turn, affects investor stock demands, which has price effects. It is this risk-sharing based endowment mechanism that allows our model to simultaneously affect the three asset pricing puzzles mentioned in the introduction and detailed in Section \ref{sec:app} below. 

While the intuition behind the risk-sharing based endowment mechanism is simple --- i.e., it is costly to rebalance to efficient positions given price-impact --- our main technical contribution gives the existence of a tractable continuous-time incomplete price-impact equilibrium model. There are two key ingredients in its construction: Exponential utilities (which could be heterogenous as in Section \ref{subsect_inhomo} below) and price-impact perceptions as in Almgren (2003).  Exponential utilities --- while not common in the standard general equilibrium asset pricing literature (which uses power utility or Epstein-Zin preferences) ---  are widely used in equilibrium models of trading such as, e.g., Grossman and Stiglitz (1980) and Vayanos (1999). Since our model requires market-clearing by heterogenous investors (due to their heterogenous stock holdings), exponential utilities make market-clearing tractable. The second ingredient, perceived price-impact, necessitates, for tractability, that we restrict investors to use trading-rate processes, which, although less common than other continuous-time processes, have been used in other equilibrium trading models including Back, Cao, and Willard (2000), Brunnermeier and Pedersen (2005), and G\^arleanu and Pedersen (2016). 

Our model's time horizon is finite but can be arbitrary long. Because of slow trading due to price-impact, our investors' heterogenous stock holdings converge gradually over time to the Radner allocations over the time horizon. Consequently, our model is non-stationary, and, in particular, the asset pricing effects of price-impact are short-term in nature. However, to the extent that investors are repeatedly shocked away from efficient risk sharing and need to trade (as in, e.g., Vayanos, 1999), the model and its asset pricing effects could be made stationary.

\section{Price-impact equilibrium}\label{Sec:2}
The definition of a Nash equilibrium in our setting is as follows:

\begin{definition}[Nash  equilibrium]\label{Nash} Continuous functions of time $A_0,...,A_3:[0,T]\to\R$ constitute a \emph{Nash equilibrium} if:

\begin{itemize}
\item[(i)] The solution $(\hat{c}_{i,t},\hat{\theta}'_{i,t})$ to trader $i$'s individual optimization problem
\eqref{optproblem} with the price-impact function \eqref{optimal101} exists for all $i\in\{1,...,I\}$.

\item[(ii)] The stock-price processes resulting from inserting trader $i$'s optimizer $\hat{\theta}'_{i,t}$ into the price-impact function $S_{i,t}$ in \eqref{optimal101} are identical for all traders $i\in\{1,...,I\}$. This common stock-price process, denoted by $\hat{S}_t$, satisfies the terminal dividend restriction \eqref{pp10}.

\item[(iii)] The individual orders $(\hat{\theta}'_{i,t})_{i=1}^I$  and corresponding holding processes $(\hat{\theta}_{i,t})_{i=1}^I$ satisfy the consistency requirement \eqref{optimal100b}.

\item[(iv)] The real-good consumption market clearing and the stock-market clearing conditions \eqref{clearing100} hold at all times $t\in[0,T]$.

\end{itemize}
$\endproof$
\end{definition}

Our main existence equilibrium existence result is based on the following technical lemma (the proof is in Appendix \ref{app:pf} below). It guarantees the existence of a solution to an autonomous  forward-backward system of coupled ODEs with forward component $\psi$ and backward components $(F,Q,Q_2,Q_{22})$. Similar forward-backward systems have appeared in equilibrium theory. For example, in Kyle (1985), the forward component is the filter and the backward components are the value function coefficients.

\begin{lemma}\label{lem:Nash} For all $\alpha>0$, there exists a constant $w\ge \frac{L^2}I$ such that the unique solutions of the coupled ODE system
\begin{footnotesize}
\begin{align}
\psi'(t) &=2\frac{F(t) Q_{22}(t)}{\alpha} \left(\psi(t)-\frac{L^2}{I}\right),\quad \psi(T)=w,\label{ODE_psi}
\\
F'(t)&= F(t)\Big(\delta -\tfrac{a^2 \sigma_D^2 }{2I}\psi(t)-\tfrac{a \left(a I \sigma_Y^2+2 a L \rho  \sigma_D \sigma_Y-2 I \mu_Y-2 L \mu_D\right)}{2 I}\Big)-1,\quad F(T)=1,\label{ODE_F}\\
Q'(t)&= -\frac{\delta }{a}+\frac{a Q(t)-\log \left(\frac{1}{F(t)}\right)+1}{a F(t)}+\frac{a \sigma_Y^2}{2}-\frac{L^2 F(t) Q_{22}(t)^2}{\alpha I^2}-\mu_Y,\quad Q(T)=0,\label{ODE_Q}
\\
Q_{2}'(t)&= a \rho  \sigma_D \sigma_Y+\frac{2 L F(t) Q_{22}(t)^2}{\alpha I}+\frac{Q_{2}(t)}{F(t)}-\mu_D,\quad Q_2(T)=0,\label{ODE_Q2}\\
Q_{22}'(t)&= a \sigma_D^2-\frac{2 F(t) Q_{22}(t)^2}{\alpha}+\frac{Q_{22}(t)}{F(t)},\quad Q_{22}(T)=0,\label{ODE_Q22}
\end{align}
\end{footnotesize}satisfy $\psi(0)=\sum_{i=1}^I\theta_{i,0}^2$. 
\end{lemma}

Next, we give our main theoretical result. In this theorem, the  parameter $\alpha>0$ is a free input parameter, which controls the temporary price-impact effect (see \eqref{optimal101B} below).  In Appendix \ref{app:cal}, we use calibrate $\alpha$ to match observed data.

\begin{theorem}\label{thm:Nash}  Let $(\psi,F,Q,Q_2,Q_{22})$ be as in Lemma \ref{lem:Nash} for initial stock endowments $\sum_{i=1}^I \theta_{i,0}=L$. A Nash equilibrium then exists in which:
\begin{itemize}
\item[(i)] The perceived investor response coefficients in \eqref{optimal100} are
\begin{align}
A_0(t)&:= \frac{I L Q_{22}(t)}{\alpha (I-1) \big(I Q_{2}(t)+2 L Q_{22}(t)\big)},\\
A_1(t)&:=\frac{A_0(t) (I-1) F(t) \big(I Q_{2}(t)+2 L Q_{22}(t)\big)}{I L},\\
A_2(t)&:= \frac{A_0(t) F(t) \big(I Q_{2}(t)-(I-2) L Q_{22}(t)\big)}{I L},\\
A_3(t)&:= A_0(t) \alpha+\frac{1}{1-I},
\end{align}
which simplifies the perceived price-impact model \eqref{optimal101} to
\begin{align}\label{optimal101B}
\begin{split}
S_{i,t}&=
F(t)D_t+F(t) \left(\frac{2 L Q_{22}(t)}{I}+Q_2(t)\right)-F(t)Q_{22}(t)\theta_{i,t}+\alpha \theta_{i,t}'.
\end{split}
\end{align}
\item[(ii)] The equilibrium interest rate $r(t)$ is given by
\begin{align}\label{r_eq}
r(t)=\delta -\frac{a^2 \sigma_D^2 }{2I}\psi(t)-\frac{a \left(a I \sigma_Y^2+2 a L \rho  \sigma_D \sigma_Y-2 I \mu_Y-2 L \mu_D\right)}{2 I}.
\end{align}
\item[(iii)] The equilibrium stock-price process is
\begin{align}\label{optimal1011}
\hat{S}_t &=F(t)D_t+F(t) \left(\frac{L Q_{22}(t)}{I}+Q_{2}(t)\right),
\end{align}
where $F(t)$ is the annuity in \eqref{ODE_F} with explicit solution \eqref{ann}.

\item[(iv)] For $i\in\{1,...,I\}$, trader $i$'s optimal order and consumption rates are:
\begin{align}
\hat{\theta}'_{i,t}&=\gamma(t)\big(\hat{\theta}_{i,t}-\frac{L}I\big),\quad \gamma(t):=\frac{F(t) Q_{22}(t)}{\alpha},\label{KLthetahat}\\
\hat{c}_{i,t} &=\tfrac{\log \big(F(t)\big)}{a}+D_t\hat{\theta}_{i,t} +\tfrac{\hat{M}_{i,t}}{F(t)}+Q(t)+\hat{\theta}_{i,t} Q_2(t)+\frac{1}{2}\hat{\theta}_{i,t}^2 Q_{22}(t)+Y_{i,t}.\label{KLchat}
\end{align}
\end{itemize}
 
\end{theorem}

\begin{remark}\label{RmkMain}

\begin{enumerate}\

\item The equilibrium stock-price process \eqref{optimal1011} is Gaussian. Such Bachelier stock-price models are common equilibrium prices in many settings including Kyle (1985), Grossman and Stiglitz (1980), and Hellwig (1980).
 
\item Our Nash equilibrium model with price-impact has stock endowment effects because the equilibrium stock holdings $\hat{\theta}_{i,t}$ for trader $i$ in \eqref{KLthetahat} 
depend on the initial endowed holdings $\theta_{i,0}$:
\begin{align} \label{explicitholdings} 
\hat{\theta}_{i,t}= \frac{L}I +\Big(\theta_{i,0}-\frac{L}I\Big) e^{\int_0^t\gamma(s)ds},\quad t\in[0,T].
\end{align}
In contrast, in the competitive Radner equilibrium (with no price-impact), trader $i$'s time $t\in(0,1]$ equilibrium holdings are $L/I$ regardless of trader $i$'s endowed holdings $\theta_{i,0}$. Section \ref{sec:app} below shows that the stock-endowment dependency ultimately allows our Nash equilibrium model to simultaneously resolve some asset pricing puzzles.

\item Heterogeneity in initial stock holdings leads to distortions in risk-sharing over time that affect asset pricing. Appendix \ref{app:pf} shows that the solution of \eqref{ODE_psi} satisfies $\psi(t)=\sum_{i=1}^I \hat{\theta}_{i,t}^2$, which is our metric for stock-holding heterogeneity.  If the initial stock endowments are equal with $\theta_{i,0}=\frac{L}I$, then $\psi'(t)=0$ from  \eqref{ODE_psi}; and hence, $\psi(t)=\frac{L^2}{I}$ for all $t\in[0,T]$. In this case, the equilibrium interest rate in \eqref{r_eq} becomes the analogous competitive Radner equilibrium interest rate given by (see Appendix \ref{ssecC} below):
\begin{align}\label{r_eq2}
r^\text{Radner}:=\delta -\frac{a^2 \sigma_D^2 L^2}{2I^2}-\frac{a \left(a I \sigma_Y^2+2 a L \rho  \sigma_D \sigma_Y-2 I \mu_Y-2 L \mu_D\right)}{2 I}.
\end{align}
For non-equal endowments (i.e., non-Pareto efficient), Cauchy-Schwart's inequality produces $\sum_{i=1}^I\theta^2_{i,0} > \frac{L^2}I$, which leads to $\psi(t)=\sum_{i=1}^I\hat{\theta}^2_{i,t} > \frac{L^2}I$ for all $t\in[0,T]$. In that case, the Nash equilibrium interest rate \eqref{r_eq} is strictly smaller than the competitive Radner equilibrium interest rate in \eqref{r_eq2}. Thus, inequality in investor stock endowments as measured by $\sum_{i=1}^{I} \hat \theta_{i,t}^2-\frac{L^2}I$ is a key factor in our model's ability to resolve the interest rate puzzle and, as shown in Section \ref{sec:app} below, also affects the other asset pricing puzzles. However, over time, the equilibrium holdings in 
\eqref{explicitholdings}  converge to equal holdings (Pareto efficient) and so these asset pricing effects are temporary.

 \item Even if the analogous competitive Radner equilibrium is Pareto-efficient (i.e., if investor income is spanned), our Nash equilibrium can be non-Pareto efficient. To see this, set $\rho^2= 1$ in the income dynamics \eqref{dYi}, which makes  the analogous  competitive Radner model complete. In this case, the interest rate \eqref{r_eq2} in the  competitive Radner equilibrium  agrees with the Pareto efficient interest rate given in \eqref{repagent2} in Appendix \ref{ssecR} below. However, as long as $\theta_{i,0}\neq \frac{L}I$ for some trader $i$, we have $\sum_{i=1}^I \theta_{i,0}^2> \frac{L^2}I$ by Cauchy-Schwartz's inequality and, consequently,  $\psi(t)=\sum_{i=1}^I \hat{\theta}_{i,t}^2> \frac{L^2}I$ by \eqref{ODE_psi}. Thus, even if the competitive Radner equilibrium is Pareto-efficient because $\rho^2= 1$ in \eqref{dYi}, the Nash equilibrium interest rate \eqref{r_eq} is strictly smaller than the Pareto-efficient equilibrium interest rate \eqref{repagent2} whenever $\theta_{i,0}\neq \frac{L}I$ for some trader $i$.

\item Unspanned investor-income randomness also affects risk-sharing and asset pricing.   Individual investor income $Y_{i,t}$  is optimally consumed, as seen in \eqref{KLchat}, and, thus,  income shocks do not directly affect optimal investor holdings. As a result, investor trading in \eqref{KLthetahat} is deterministic, which simplifies the modeling of the stock endowment effects.  However, the parameters of the investor income process do affect asset pricing in \eqref{r_eq} and \eqref{optimal1011} and the optimal trading rate $\hat{\theta}'_{i,t}$ in  \eqref{KLthetahat}. Thus, imperfect risk-sharing due to both distortions in initial stock endowments and unspanned (idiosyncratic) shocks to investor income has asset-pricing effects with price-impact.  

\item The proof of Theorem \ref{thm:Nash} in  Appendix \ref{app:pf} is based on the standard dynamical programming principle and HJB equations. Thus, by definition, the individual optimization problems in our Nash equilibrium model are time-consistent. However, it might appear that our Nash equilibrium model is time-inconsistent given that the optimal holdings $\hat{\theta}_{i,t}$ in \eqref{explicitholdings} depend on the endowed holdings $\theta_{i,0}$  (see, e.g., the discussion in Remark 3 on p.455 in Basak, 1997).  The explanation for why our Nash equilibrium is time-consistent while the equilibrium holdings $\hat{\theta}_{i,t}$ depend on the endowed holdings $\theta_{i,0}$ lies in the state-processes and controls used in the proof of Theorem \ref{thm:Nash}, summarized in Table \ref{tab0}:

\begin{table}[ht]
\centering  \caption{State-processes and controls used.}
\label{tab0}
\vspace{.2cm}
\begin{tabular}{c|c|c}
\hline\hline
  & State processes& Controls\\
   \hline
Nash&$ M_{i,t},D_t,\theta_{i,t}$& $c_{i,t},\theta_{i,t}' $\\
Radner and Pareto&$X_{i,t}$ & $c_{i,t},\theta_{i,t}$ \\
\hline\hline
\end{tabular}
\end{table}
For time-consistent optimization problems, the initial control values cannot appear in the optimal controls. However, because the trading rate $\theta'_{i,t}$ is the control --- not stock holdings $\theta_{i,t}$ --- in the Nash equilibrium model, the endowment $\theta_{i,0}$ can (and do) appear in the time-consistent individual optimal holdings $\hat{\theta}_{i,t}$  in \eqref{explicitholdings}. Likewise, the Radner and Pareto equilibrium models are time-consistent, and so the endowment $\theta_{i,0}$ cannot (and do not) appear in the individual optimal holdings $\frac{L}I$.

\end{enumerate}
\end{remark}

\section{Asset-pricing puzzles}\label{sec:app}

This section shows that our continuous-time price-impact equilibrium model  produces material differences relative to the analogous Pareto-efficient equilibrium. In particular,  based on the C-CAPM from Breeden (1979), Appendix \ref{ssecR} derives the analogous Pareto-efficient equilibrium where all investors act as price-takers and markets are complete. We show how price-impact simultaneously affects the three main asset-pricing puzzles (risk-free rate, equity premium, and volatility). We do this both analytically and by illustrating the equilibrium differences in a numerical example. The differences between our model and the Pareto-efficient equilibrium are due to perceived price-impact, heterogenous stock holdings, and  market incompleteness (due to idiosyncratic income risk when $\rho^2< 1$).

Our conclusion is that, by using the Pareto-efficient equilibrium model as a benchmark, our price-impact Nash equilibrium model can simultaneously help resolve the risk-free rate puzzle of Weil (1989), and the volatility puzzle of LeRoy and Porter (1981) and Shiller (1981). Price-impact also moves the Sharpe ratio in the right direction qualitatively for the equity premium puzzle of Mehra and Prescott (1985), but the effect is quantitatively small. These empirical works on asset-pricing puzzles compare a competitive representative agent model with historical data. Such representative agent models are (effectively) complete and therefore also Pareto efficient by the First Welfare theorem. Therefore, we use the Pareto efficient equilibrium interest rate and stock-price process as benchmarks.

\subsection{Discussion}

First, consider the risk-free rate puzzle of Weil (1989). Pareto-efficient  equilibrium models predict interest rates that are too high compared to empirical evidence. For the Nash equilibrium interest rate $r(t)$ in  \eqref{r_eq}, the analogous competitive interest rate $r^\text{Radner}$ in  \eqref{r_eq2}, and the analogous Pareto-efficient interest rate $r^\text{Pareto}$  in \eqref{repagent2} in Appendix \ref{ssecR}, we have the ordering
\begin{align}\label{rodering}
r(t) \le r^\text{Radner} \le r^\text{Pareto},\quad t\in[0,T].
\end{align}
Whenever there is unspanned income risk (i.e., when $\rho^2 <1$), Christensen, Larsen, and Munk (2012) show that $r^\text{Radner} <r^\text{Pareto}$ due to a precautionary saving effect. Here, we find $r(t) < r^\text{Radner}$ whenever there is stock-endowment inequality in that $\theta_{i,0}\neq L/I$ for some trader $i\in \{1,...,I\}$.\footnote{Even without idiosyncratic income risks (i.e., $\rho^2 =1$ so that $r^\text{Radner} =r^\text{Pareto}$), we have $r(t) < r^\text{Radner}$.} The intuition is that price-impact costs cause investors to rebalance more slowly, which exacerbates risk-bearing inefficiency, which, in turn, magnifies stock risk and increases bond demand.

Second, consider the volatility puzzle of LeRoy and Porter (1981) and Shiller (1981). Pareto-efficient  models predict a stock-price volatility that is too low compared to empirical evidence. The ordering \eqref{rodering} reverses the annuity ordering:
\begin{align}\label{Fodering}
F(t) \ge F^\text{Radner}(t) \ge F^\text{Pareto}(t),
\end{align}
where $F(t)$ is given by the ODE \eqref{ODE_F} and
\begin{align}
\frac{d}{dt}F^\text{Radner}(t)&= F^\text{Radner}(t) r^\text{Radner}-1,\quad F^\text{Radner}(T)=1,\label{ODE_FRadner}\\
\frac{d}{dt}F^\text{Pareto}(t)&= F^\text{Pareto}(t) r^\text{Pareto}-1,\quad F^\text{Pareto}(T)=1.\label{ODE_FPareto}
\end{align}
Consequently, the ordering \eqref{Fodering} and the equilibrium stock-price processes \eqref{optimal1011}, \eqref{SRad}, and \eqref{repagentS} produce the volatility ordering measured by quadratic variation
\begin{align}\label{volodering}
d\langle \hat{S}\rangle_t \ge d\langle S^\text{Radner}\rangle_t\ge d\langle S^\text{Pareto}\rangle_t,
\end{align}
with strict inequalities whenever the inequalities in \eqref{rodering} are strict. The intuition is that the multiplication of the current dividend $D_t$ by $F(t)$ in \eqref{optimal1011} represents an annuity-valuation effect for the stream of future dividends following $D_t$ at time $t$. Thus, lower interest rates in the Nash equilibrium intensify this annuity effect. 

Third, consider the equity premium puzzle of Mehra and Prescott (1985). Pareto-efficient  models predict the stock's excess return over the risk-free rate to be too low compared to empirical evidence. To address the equity premium puzzle, we start by recalling the definition of the equity premium:
\begin{align}\label{EP}
\begin{split}
\text{EP}(t) :&= \E\left[\frac{\hat{S}_t-\hat{S}_0 +\int_0^t D_u e^{\int_u^tr(s)ds}du}{\hat{S}_0}\right]-\Big(e^{\int_0^t r(u)du}-1\Big),\quad t\in[0,T].
\end{split}
\end{align}
In \eqref{EP}, the interest rate $r(t)$ is given in \eqref{r_eq} with the corresponding (deterministic) money market account price process is $e^{\int_0^t r(u)du}$ and the equilibrium stock-price process $\hat{S}_t$ is given in \eqref{optimal1011}. Based on \eqref{EP}, we define the Sharpe ratio measured over a time interval $[0,t]$ as
\begin{align}\label{DSR}
\begin{split}
\text{SR}(t) :=& \frac{\text{EP}(t)}{\V\Big[\frac{\hat{S}_t-\hat{S}_0+\int_0^t D_u\frac{S_t^{(0)}}{S_u^{(0)}} du}{\hat{S}_0}-\big(e^{\int_0^t r(u)du}-1\big)\Big]^\frac12}, \quad t\in(0,T],
\end{split}
\end{align}
where $\V[\cdot]$ in the denominator in \eqref{DSR} is the variance operator. Because models based on noise generated by Brownian motions produce expected returns and variances growing linear in $t$ for $t>0$ small, we consider the time-normalized Sharpe ratio defined by $\frac{\text{SR}(t)}{\sqrt{t}}$ for a horizon $t\in(0,T]$. 
The \emph{instantaneous} Sharpe ratio is defined as the limit
\begin{align}\label{instantSR}
\begin{split}
\lambda:&=\lim_{t\downarrow 0}\frac{\text{SR}(t)}{\sqrt{t}}\\
&=\frac{a}I (L\sigma_D+I\sigma_Y\rho).
\end{split}
\end{align}
The coefficient $\lambda$ in \eqref{instantSR} is called the \emph{market price of risk} because the dynamics of the Nash equilibrium stock-price process \eqref{optimal1011} are
\begin{align}
\begin{split}
d\hat{S}_t &=F(t)\Big(dD_t+\frac{L Q'_{22}(t)}{I}dt+Q'_{2}(t)dt\Big) + \frac{F'(t)}{F(t)} \hat{S}_tdt\\
&=\Big(r(t)\hat{S}_t - D_t +\frac{a\sigma_D(L\sigma_D +I\rho\sigma_Y)F(t)}I \Big)dt + F(t) \sigma_D dB_t\\
&=\big(r(t)\hat{S}_t - D_t \big)dt + F(t) \sigma_D \big(dB_t+\lambda dt\big).
\end{split}
\end{align}
The Sharpe ratios SR$^\text{Radner}(t)$ and SR$^\text{Pareto}(t)$ are defined analogously for the competitive Radner and Pareto-efficient equilibrium stock-price processes $S^\text{Radner}$ from \eqref{SRad}  and $S^\text{Pareto}$ from  \eqref{repagentS} and interest rates $r^\text{Radner}$ from \eqref{r_eq2} and  $r^\text{Pareto}$ from \eqref{repagent2}. 

The numerics in the next section  illustrate that our Nash equilibrium model with price-impact can produce a higher Sharpe ratio than both the Radner and Pareto equilibrium models; that is, we shall see 
\begin{align}\label{orderSR}
\text{SR}(t)\ge \text{SR}^\text{Radner}(t)\ge \text{SR}^\text{Pareto}(t),\quad t\in[0,T],
\end{align}
for reasonable model parameters. Because the equity premium puzzle involves empirical Sharpe ratios estimated over discrete horizons (e.g., monthly or annually), the ordering of finite-horizon $[0,t]$ Sharpe ratios in \eqref{orderSR} is the relevant measure. However, Section \ref{sec:num} below shows that the magnitudes of the Sharpe ratio difference in \eqref{orderSR} are quantitatively small.  The reason is that, as $t \downarrow 0$, the time-normalized discrete-horizon Sharpe ratios in all three models (Nash, Radner, and Pareto) are anchored to the same instantaneous Sharpe ratio $\lambda$ in \eqref{instantSR}.\footnote{There are several ways model incompleteness can produce a different instantaneous Sharpe ratio in the competitive (i.e., price-taking) Radner equilibrium model relative to the analogous efficient Pareto model: (i) Traders can be restricted to only consume discretely as in Constantinides and Duffie (1996), (ii) The underlying filtration can have jumps as in, e.g., Barro (2006) and Larsen and Sae-Sue (2016), and (iii) Non-time additive utilities as in, e.g., Bansal and Yaron (2004). Christensen and Larsen (2014, p.273) prove that the instantaneous Sharpe ratios in the Radner and Pareto equilibrium models always agree in a setting based on exponential utilities in a continuous-time consumption model with noise generated by Brownian motions. See also the discussion in Cochrane (2008, p.310).}
Note here that the Sharpe ratio \eqref{DSR} is a ratio of integrals and not an integral of instantaneous Sharpe ratios. Therefore, for $t>0$ small, the Nash Sharpe ratios \eqref{DSR} are similar to the analogous Sharpe ratios in the Radner and Pareto equilibria. However, over longer horizons $t>0$, our Nash equilibrium model with price-impact can produce modestly bigger Sharpe ratios \eqref{DSR} than the analogous Radner and Pareto equilibrium models.

\subsection{Numerics}\label{sec:num}

This section presents calibrated numerics to illustrate the effect of price-impact on all three asset-pricing puzzles. In our numerics, time is measured on an annual basis (i.e., one year is $t=1$). We normalize the outstanding stock supply to $L:=100$.  As noted in Remark \ref{RmkMain}.3, the key quantity in explaining the asset pricing puzzles is the heterogeneity in investors' initial endowments as measured by the difference $\sum_{i=1}^I\theta_{i,0}^2- \frac{L^2}I\ge0$ which is a metric for the distance of the initial stock endowments from Pareto efficiency. To provide some intuition for this difference, we note that the cross-sectional average and standard deviation of a set of initial stock endowments $\vec{\theta}_0:=\{\theta_{1,0},...,\theta_{I,0}\}$ are
\begin{align}\label{mean_SD}
\begin{split}
\text{mean}\big[\vec{\theta}_{0}\big] :&= \frac1I \sum_{i=1}^I\theta_{i,0}\\&=\frac{L}{I},\\
\text{SD}\big[\vec{\theta}_{0}\big] :&= \sqrt{\frac1I \Big(\sum_{i=1}^I\theta^2_{i,0}-\frac{L^2}I\Big)}\\
& = \sqrt{\frac1I \big(\psi(0)-\frac{L^2}I\big)},
\end{split}
\end{align}
where $\psi(t)$ is the function from \eqref{ODE_psi}.

The utility parameters for \eqref{optproblem} in our numerics are
\begin{align}\label{par1}
 \delta :=0.02,\quad a:=2.
\end{align}
The annual time-preference rate $\delta$ is consistent with calibrated time preferences in Bansal and Yaron (2004), and the level of absolute risk aversion $a$ is from the numerics in Christensen, Larsen, and Munk (2012).  The coefficients for the arithmetic Brownian motion for the stock  dividends in \eqref{dD} are
\begin{align}\label{par11}
\begin{split}
\mu_D := 0.0201672,\quad  \sigma_D:= 0.0226743,\quad D_0:=1.
\end{split}
\end{align}
The parameterizations of $\mu_D$ and $\sigma_D$ are the annualized mean
and standard deviation of monthly percentage changes in aggregate real US stock market dividends from January 1970 through December 2019 (from Robert Shiller's website \url{http://www.econ.yale.edu/~shiller/data.htm}). The starting dividend rate $D_0=1$ in \eqref{par11} is a normalization. The annualized income volatility and income-dividend correlation are from the numerics in Christensen, Larsen, and Munk (2012):
\begin{align}\label{par110}
 \sigma_Y=0.1,\quad \rho=0.
\end{align} 
The drift $\mu_Y$ and number of investors $I\in\N$ are found by calibrating the Radner equilibrium model so that
\begin{align}
\lambda = 0.302324,\quad r^\text{Radner} = 8.137\%,
\end{align}
which produces the remaining coefficients\footnote{The discount rate $\delta$, dividend parameters $\mu_D$ and $\sigma_D$, and income parameters $\mu_Y$ and $\sigma_Y$ are all quoted in decimal form where 0.01 = 1\%.}
\begin{align}\label{par111}
\mu_Y:=-0.0709146,\quad I=15.
\end{align}
We set the model horizon $T$ to $T := 3$ years. In our analysis we found that  our numerics are relatively insensitive to $T$ once $T$ is sufficiently large.

\begin{figure}[!h]
\begin{center}
\caption{Trajectories of interest rates (Plots A and B),  stock-price volatility (Plots C and D), and Sharpe ratio differences SR$(t)-$ SR$^\text{Radner}(t)$ (Plots E and F) for $t\in[0,1]$ over the first year. The model parameters are given in \eqref{par1}, \eqref{par11}, \eqref{par110}, \eqref{par111}, $L:=100$,  $T:=3$ years, and the time discretization uses 250,000 rounds of trading per year. }
\begin{footnotesize}
$\begin{array}{cc}
\includegraphics[width=6cm, height=4.5cm]{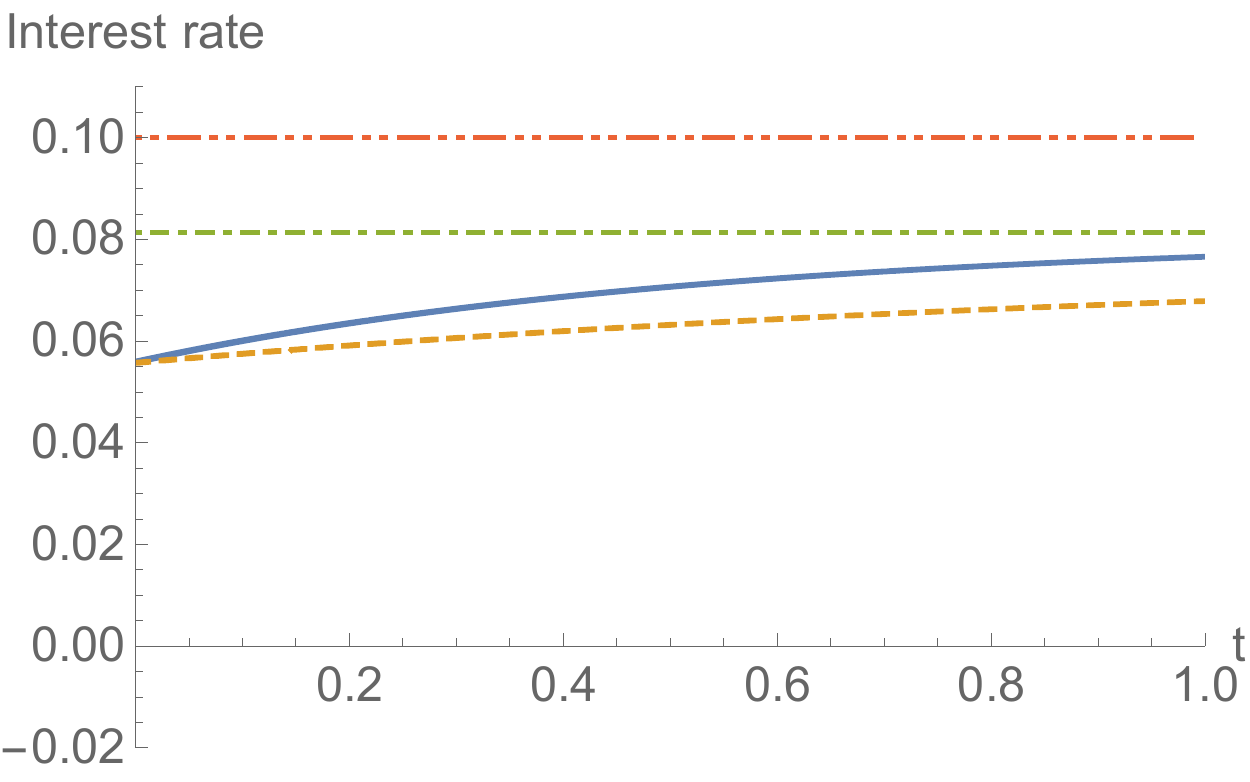} &
\includegraphics[width=6cm, height=4.5cm]{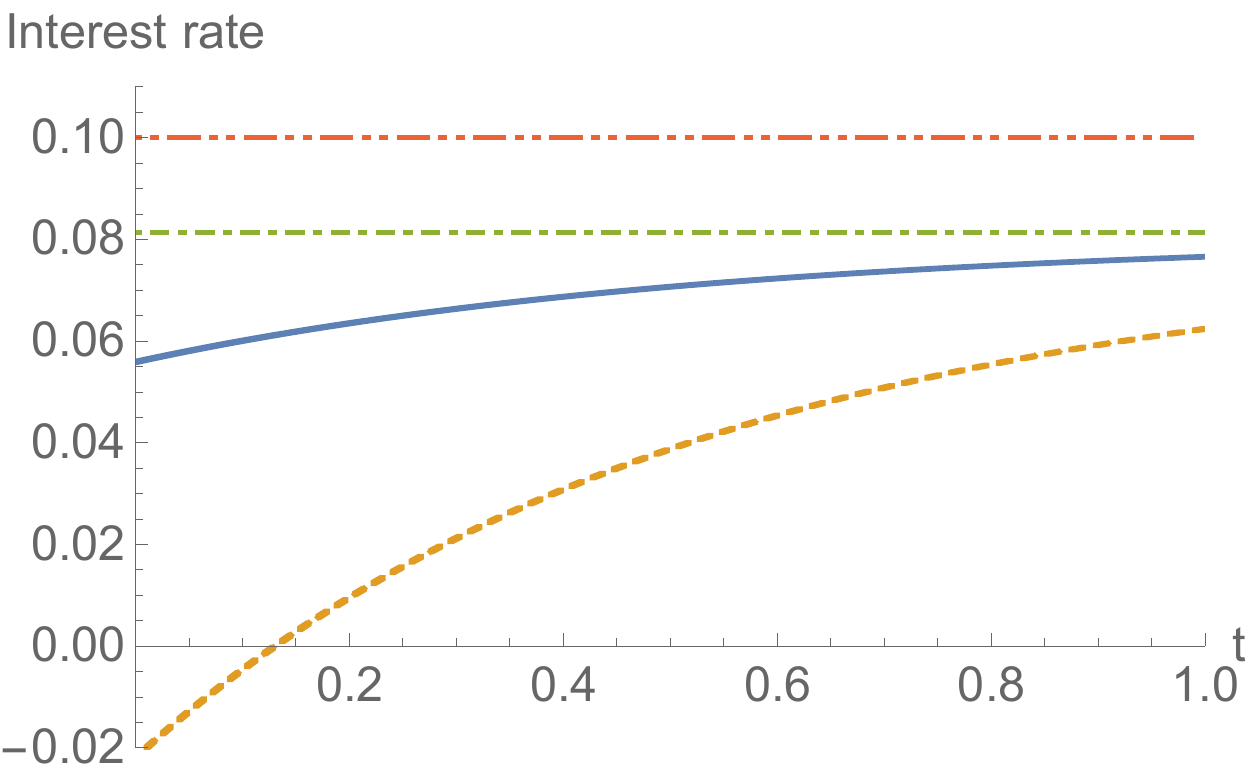}
\\ 
\text{A: } \text{SD}\big[\vec{\theta}_{0}\big]:= 5& \text{B: } \alpha:=0.002\\
 \text{Nash: $\alpha:=0.002$ (-----)},\; \alpha:=0.01\;(- -), &\text{Nash $\text{SD}\big[\vec{\theta}_{0}\big]:= 5$  (-----)},\; \text{SD}\big[\vec{\theta}_0\big]:= 10 \;(- -),\\
 \text{Radner: ($-\cdot-$)},\;\text{Pareto: ($-\cdot\cdot-$)} & \text{Radner: ($-\cdot-$)},\;\text{Pareto: ($-\cdot\cdot-$)}\\ 
\includegraphics[width=6cm, height=4.5cm]{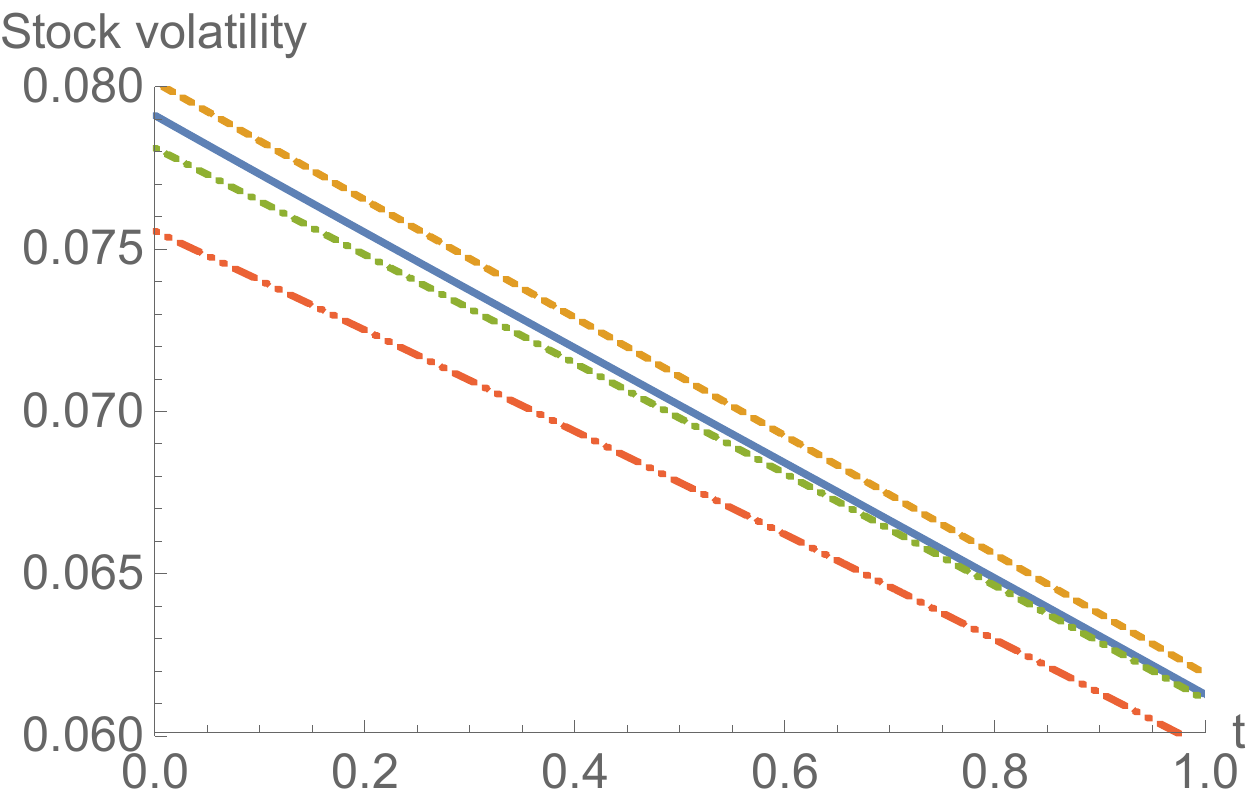} &\includegraphics[width=6cm, height=4.5cm]{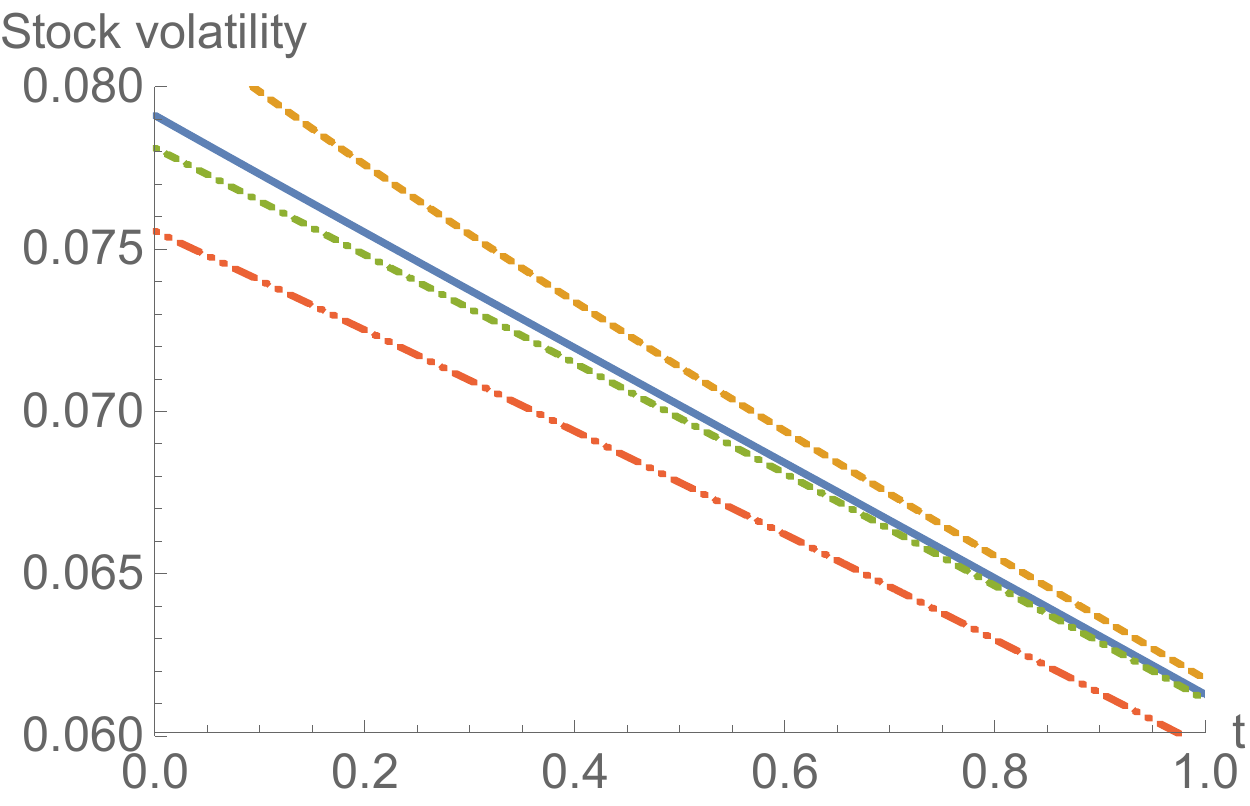}
\\ 
\text{C: } \text{SD}\big[\vec{\theta}_{0}\big]:= 5& \text{D: } \alpha:=0.002\\
 \text{Nash: $\alpha:=0.002$ (-----)},\; \alpha:=0.01\;(- -), &\text{Nash: $\text{SD}\big[\vec{\theta}_{0}\big]:= 5$  (-----)},\; \big[\vec{\theta}_0\big]:= 10 \;(- -),\\
 \text{Radner: ($-\cdot-$)},\;\text{Pareto: ($-\cdot\cdot-$)} & \text{Radner: ($-\cdot-$)},\;\text{Pareto: ($-\cdot\cdot-$)}\\ 
\includegraphics[width=6cm, height=4.5cm]{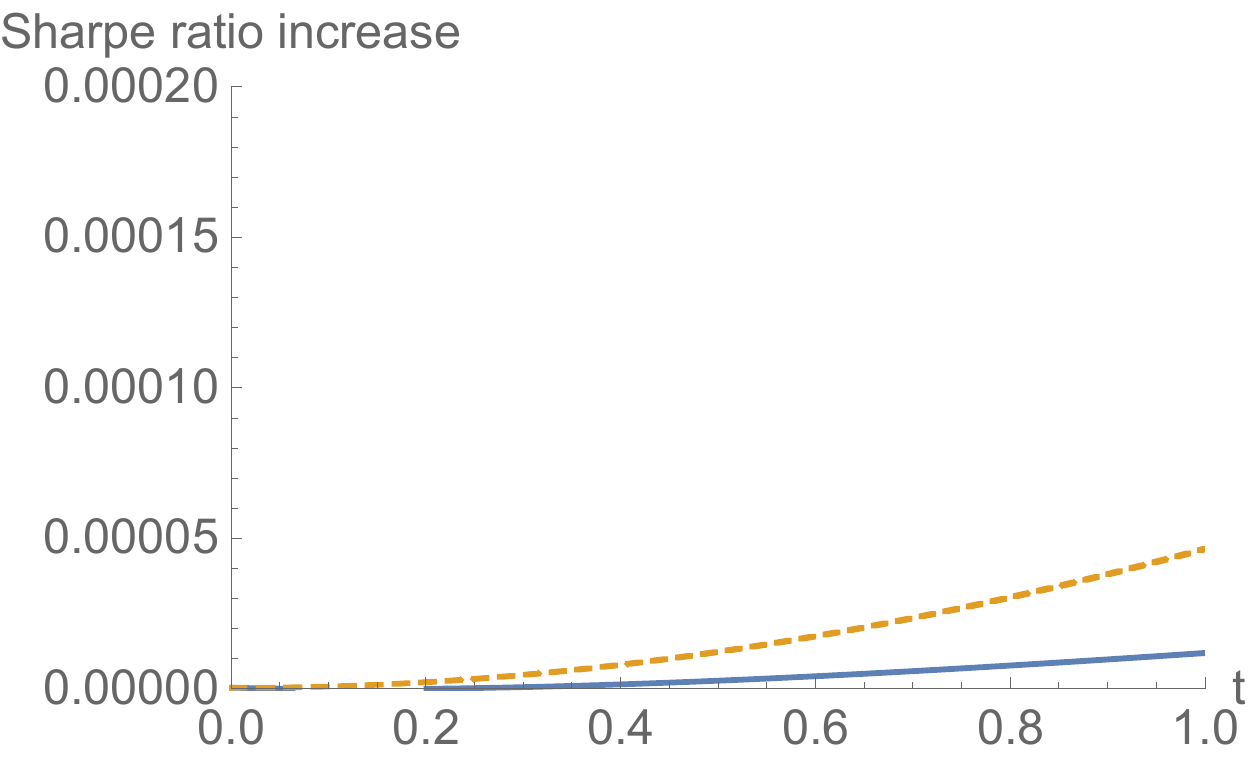} 
&\includegraphics[width=6cm, height=4.5cm]{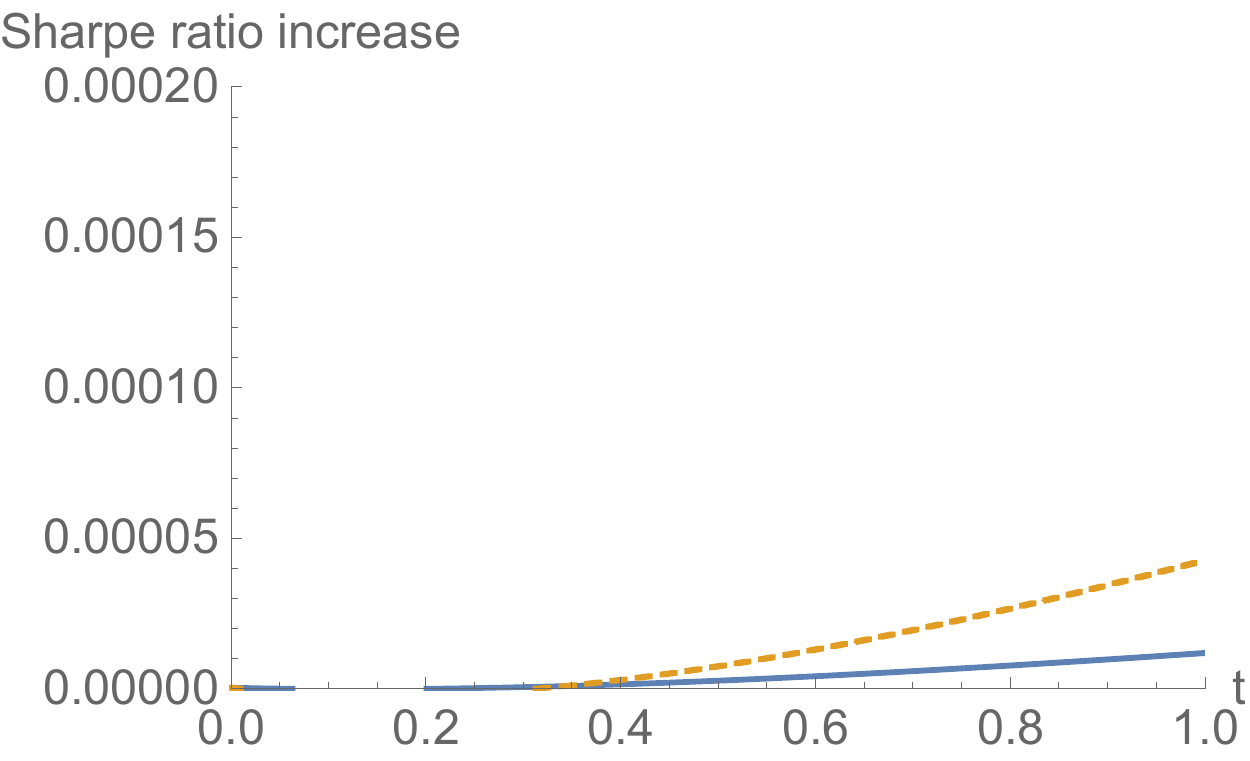}
\\ 
\text{E: } \text{SD}\big[\vec{\theta}_{0}\big]:= 5& \text{F: } \alpha:=0.002\\
 \text{Nash: $\alpha:=0.002$ (-----)},\; \alpha:=0.01\;(- -) &\text{Nash: $\text{SD}\big[\vec{\theta}_{0}\big]:= 5$ (-----) },\; \text{SD}\big[\vec{\theta}_0\big]:= 10 \;(- -)
\end{array}$
 \end{footnotesize}
\label{fig1}
\end{center}
\end{figure}

We illustrate that price-impact in the Nash equilibrium can have a material effect on asset pricing relative to the analogous Pareto-efficient equilibrium. Figure \ref{fig1} shows interest-rate and stock return-volatility trajectories over a year $t\in [0,1]$ for the Nash equilibrium with price-impact, the price-taking Radner equilibrium, and the corresponding Pareto-efficient equilibrium.  For visibility, Figure \ref{fig1} also shows differences in Sharpe ratios between the Nash and Radner equilibria  since these numerical values are small.

The Nash model with price-impact has two additional parameters relative to the competitive Radner model: The transitory price-impact coefficient $\alpha$ in \eqref{optimal101B} and the difference $\sum_{i=1}^I\hat{\theta}_{i,t}^2 - L^2/I= \psi(t) - L^2/I$ for deviations of initial stock endowments from the equal stock holdings, which is related to the SD[$\vec{\theta}]$ in \eqref{mean_SD}.  Figure \ref{fig1} illustrates the sensitivity of asset pricing moments to these two parameters.

Figure \ref{fig1}, Plots A, C, and E show the effects of varying the temporary price-impact parameter $\alpha>0$. Of course, when $\alpha > 0$ is close to zero, our Nash equilibrium is close to the Radner equilibrium. In our numerics, we consider two transitory price-impact parameters of $\alpha \in \{0.01,0.002\}$. Appendix \ref{app:cal} shows that $\alpha=0.002$ is roughly consistent with transitory price-impact estimates in Almgren et al. (2005). To put them in perspective, a price-impact of $\alpha = 0.002$ means if an investor trades at a constant rate $\theta'_{i}=265$ to sell $\int_0^{\frac1{265}}\theta'_idt=1$ unit of the stock over a day (i.e., a large daily parent trade of 1 percent of $L = 100$ shares outstanding), the associated transitory price increase at each time $t$ in the day would be $0.002\times 265=0.53$. Given that the stock (with $\alpha =0.002$ and SD$[\vec{\theta}]=5$) has an endogenous initial equilibrium price of $\hat S_0 = 3.5737$ (see Table \ref{tab2}), this corresponds to a sustained percentage transitory  price-impact of $\frac{0.002\times 265}{3.5737}= 14.83\%$ over the day. 

The price-impact feature in the Nash equilibrium can produce up to a 2\% annual interest rate reduction (the reduction is biggest for shorter horizons). We see that the stock-return volatility increases by around 0.25\% relative to the Radner volatility. The impact on the Sharpe ratio, while in the right direction qualitatively, is quantitatively small. The Sharpe ratio effects are biggest for longer horizons (as already discussed after \eqref{orderSR}, for short horizons the Sharpe ratios are anchored to the instantaneous Sharpe ratio $\lambda$). Finally, from Plots A, C, and E, we see that all three asset-pricing impacts are increasing in the temporary price-impact coefficient $\alpha>0$.

Figure \ref{fig1}, Plots B, D, and F consider the effect of different levels of stock-endowment inequality (SD$[\vec{\theta}_0]\in \{5,10\})$. As $\sum_{i=1}^I \theta_{i,0}^2$ approaches the lower bound $\frac{L^2}{I}$ from Cauchy-Schwartz's inequality, the Nash equilibrium converges to the Radner equilibrium. Plots B, D, and F, show that that all three asset-pricing impacts are increasing in investor heterogeneity as measured by the difference $\sum_{i=1}^I \theta_{i,0}^2-\frac{L^2}I$.

\begin{table}[ht]
\centering  \caption{Numerical output. Numbers inside () and [\,] are from the analogous (Radner) and  [Pareto-efficient] equilibria.}
\label{tab2}
\vspace{.2cm}
\begin{footnotesize}
\begin{tabular}{c|c|c|c|c|c}
\hline\hline
SD$\big[\vec{\theta}_0\big]$ &$\alpha$ & $\hat{S}_0$& $\sum_{i=1}^I\hat{\theta}_{i,T}^2$& SR(1) &$r$(0)\\
    \hline
5&0.002  & 3.5737 (3.5276) [3.4112]  &677.4 & 0.3010 & 5.558\% (8.137\%) [10.003\%] \\
10&0.002  & 3.7135 (3.5276) [3.4112]  &708.9 & 0.3010 & -2.196\% (8.137\%) [10.003\%] \\
5&0.01  & 3.6200 (3.5276) [3.4112]  &782 & 0.3011 & 5.569\% (8.137\%) [10.003\%] \\

\hline\hline
\end{tabular}
\end{footnotesize}

\end{table}

\newpage

\section{Model extensions}\label{sec:extension}

Our analysis has shown how to construct a parsimonious and tractable model of price-impact in continuous-time. However, the following three model extensions illustrate our Nash equilibrium model's analytical robustness to variations. 

\subsection{Discrete-orders }\label{discreteorders}
To illustrate that we can allow traders to also place discrete orders (i.e., block orders) as well as consumption plans with lump sums, we consider a simple case. We allow the traders to place block orders and consume in lumps at time $t=0$ after which they trade using order rates and consume using consumption rates for $t\in(0,T]$. 

First, we start with block orders and use $\theta_{i,0-}$ to denote trader $i$'s initial stock endowment so that $\Delta \theta_{i,0}:= \theta_{i,0}-\theta_{i,0-}$ denotes the block order at time $t=0$. In addition to \eqref{optimal100} for $t\in(0,T]$, we conjecture the response at time $t=0$ for trader $j\neq i$ to be
\begin{align}\label{optimal100B}
\begin{split}
\Delta\theta_{j,0} &=  \beta_0\big(F(0)D_{0}-S_{i,0}\big)+\beta_1 \theta_{j,0-}+\beta_2 \theta_{i,0-}+\beta_3\Delta \theta_{i,0},
\end{split}
\end{align}
where $(\beta_0,..,\beta_3)$ are constants (to be determined). The price-impact function trader $i$ perceives is found using the stock-market clearing condition at time $t=0$ when summing \eqref{optimal100B}:
\begin{align}\label{optimal88B}
\begin{split}
0 &=  (I-1)\beta_0 \big(F(0)D_{0}-S_{i,0}\big)+\beta_1(L-\theta_{i,0-})\\
&+(I-1)\big(\beta_2\theta_{i,0-}+\beta_3\Delta \theta_{i,0}\big) + \Delta \theta_{i,0}.
\end{split}
\end{align}
Provided that $\beta_0\neq 0$, we can solve \eqref{optimal88B} for trader $i$'s perceived stock market-clearing price at time $t=0$:
\begin{align}\label{optimal101S0}
\begin{split}
S_{i,0}&= D_0 F(0)+ \frac{\beta_1 L}{\beta_0 (I-1)} + \frac{\beta_2(I-1)-\beta_1}{\beta_0(I-1)}\theta_{i,0-}+ \frac{\beta_3 (I-1)+1}{\beta_0 (I-1)}\Delta\theta_{i,0}.
\end{split}
\end{align}

Second, we introduce time $t=0$ lump sum consumption.  Because stock prices are denoted ex dividend, the initial wealth is
\begin{align}\label{X0}
\begin{split}
X_{i,0} =(D_{0} + S_{i,0})\theta_{i,0-} +Y_{i,0}- C_{i},\quad i\in\{1,...,I\},
\end{split}
\end{align}
where $C_{i}$ is trader $i$'s lump sum consumption at time $t=0$ (to be determined). The expression for $X_{i,0}$ in \eqref{X0} follows from the normalization that all strategic traders have zero  endowments in the money market account. 
By using \eqref{X0}, the time $t=0$ money market account balance of \eqref{M} for trader $ i\in\{1,...,I\}$ is given by
\begin{align}\label{M0}
\begin{split}
M_{i,0} :=& \,X_{i,0} - S_{i,0}\theta_{i,0} \\
=&\,D_{0}\theta_{i,0-}  -S_{i,0}\Delta \theta_{i,0}  +Y_{i,0}- C_{i}.
\end{split}
\end{align}

Next, we show how to modify to  the objective in \eqref{optproblem} to allow for both time $t=0$ lump sum consumption $C_i$ and block orders $\Delta \theta_{i,0}$. Trader $i$'s optimization problem becomes:
\begin{align}\label{optproblemAconsum}
\begin{split}
&\inf_{(\Delta\theta_{i,0},C_i)\in\R^2,\,(\theta'_{i},c_{i})\in \sA}  \E\left[e^{-a C_{i}}+\int_0^T e^{-a c_{i,t} -\delta t}dt + e^{-a (X_{i,1}+Y_{i,T})-\delta T}\right]\\
&=\inf_{(\Delta\theta_{i,0},C_i)\in\R^2}\Big(e^{-a C_{i}}+v(0,M_{i,0},D_0,\theta_{i,0},Y_{i,0})\Big),
\end{split}
\end{align}
where $v$ is the value function defined below in \eqref{valn} in Appendix \ref{app:pf} corresponding to the objective in  \eqref{optproblem}. To minimize the objective in \eqref{optproblemAconsum}, we insert $M_{i,0}$ from \eqref{M0} and $\theta_{i,0}=\theta_{i,0-}+\Delta \theta_{i,0}$ into the last line in \eqref{optproblemAconsum} and minimize to produce the optimal initial block order and lump sum consumption. For example, we have
\begin{align}\label{discreteorder0}
\begin{split}
\hat{S}_0 &= F(0)\Big(D_0+\tfrac{L Q_{22}(0)}{I}+Q_2(0)\Big),\\
 \hat{\theta}_{i,0} &=
\theta_{i,0-}+\beta(\theta_{i,0-}-\frac{L}I),
\end{split}
\end{align}
where $\beta$ is a free model parameter (similar to $\alpha$ in Theorem \ref{thm:Nash}). From \eqref{discreteorder0}, we see that $\hat{S}_0$ matches the initial stock price in \eqref{optimal1011}. Moreover, because of price-impact, we also see from \eqref{discreteorder0} that  trader $i$ does not immediately jump to the Pareto efficient holdings $\frac{L}{I}$.

\subsection{Heterogenous utilities and incomes}\label{subsect_inhomo}
In addition to the $I$ traders with utilities as in \eqref{optproblem}, we introduce a second group of traders indexed by $i\in\{I+1,..., I +\bar{I}\}$, $\bar{I}\in\N$, with utilities given by
\begin{align}\label{Uibar}
-e^{-\bar{a} c -\bar{\delta}t },\quad c\in\R,\quad t\in[0,T].
\end{align}
The coefficients  $\bar{a}>0$ and $\bar{\delta}\ge0$ are potentially different from those in \eqref{optproblem}. These traders have income rate processes similar to \eqref{dYi}:
\begin{align}\label{dYibar}
dY_{i,t} := \bar{\mu}_Y dt + \bar{\sigma}_Y\big(\bar{\rho}  dB_t+\sqrt{1-\bar{\rho}^2}dW_{i,t}\big),\quad Y_{i,0} \in \R,
\end{align}
but the coefficients $(\bar{\mu}_Y,\bar{\sigma}_Y,\bar{\rho})$ are potentially different from those in  \eqref{dYi}.

In this heterogenous setting, a Nash equilibrium is given by deterministic functions of time $A_0(t),... ,A_3(t)$ and $\bar{A}_0(t),..,\bar{A}_3(t)$. The response functions for trader $i\in\{1,..., I \}$ is as in \eqref{optimal100} whereas for trader $i\in\{I+1,..., I +\bar{I}\}$ the conjectured response function for trader $j\neq i$ is
\begin{align}\label{optimal100bar}
\begin{split}
\theta'_{j,t} &:=  \bar{A}_0(t)\big(F(t)D_{t}-S_{i,t}\big)+\bar{A}_1(t)\theta_{j,t}+\bar{A}_2(t)\theta_{i,t}+\bar{A}_3(t)\theta'_{i,t}.
\end{split}
\end{align}
There are two different perceived stock-price processes with price-impact. For trader $i\in\{1,..., I \}$, the stock-price process subject to trader $i$'s choice of $\theta'_{i,t}$ is found by solving
\begin{align}\label{Sismall}
0 = \sum_{j=1,j\neq i}^I \theta'_{j,t} + \sum_{j=I+1}^{I+\bar{I}} \theta'_{j,t} + \theta'_{i,t}
\end{align}
 for $ S_{i,t}$. Similarly, for trader $i\in\{I+1,..., I+\bar{I} \}$, the stock-price process subject to trader $i$'s choice of $\theta'_{i,t}$ is found by solving
\begin{align}\label{Sismallll}
0= \sum_{j=1}^I \theta'_{j,t} + \sum_{j=I+1,j\neq i}^{I+\bar{I}} \theta'_{j,t} + \theta'_{i,t}
\end{align}
for $ S_{i,t}$. By doubling the number of $Q$ functions, the existence result in Theorem \ref{thm:Nash} can be modified to include this extension.

\subsection{Penalties}

In this section, we replace the objective \eqref{optproblem} with
\begin{align}\label{optproblempenalty}
\inf_{(\theta'_{i},c_{i})\in \sA}  \E\left[\int_0^T e^{-a c_{i,t} -\delta t}dt + e^{-a (X_{i,1}+Y_{i,T}-L_{i,T})-\delta T}\right],\quad i=1,...,I,
\end{align}
where $L_{i,T}$ is a penalty term. We consider two specifications of $L_{i}$. First, we can incorporate high-frequency traders (HFTs) who are incentivized to hold zero positions over time. We do this by defining  the penalty processes:
\begin{align}\label{Lit}
L_{i,t}:= \int_0^t \kappa(s) \theta_{i,s}^2ds,\quad t\in[0,T],\quad i=1,...,I.
\end{align}
The deterministic function $\kappa:[0,T]\to [0,\infty)$ in \eqref{Lit} is a penalty-severity function.  The strength of $\kappa(t)$ for $t\in[0,T]$ can vary periodically for times during overnight periods vs during trading days to give HFTs stronger incentive to hold no stocks overnight. Similar to the extension in subsection \ref{subsect_inhomo}, it is also possible to consider multiple groups of homogenous traders where traders in different groups have identical penalty functions but different groups can have different penalty-severity functions.

Second, we can approximate transaction costs by penalizing trading rates (as in, e.g., G\^arleanu and Pedersen 2016). We do this by defining the penalty processes:
\begin{align}\label{Lit2}
L_{i,t}:= \frac12 \lambda \int_0^t (\theta_{i,s}')^2ds,\quad t\in[0,T],\quad i=1,...,I.
\end{align}
The constant $\lambda>0$ in \eqref{Lit2} is interpreted as a transaction cost parameter.

By altering the ODEs, the existence result in Theorem \ref{thm:Nash} can be modified to include both penalties \eqref{Lit} and \eqref{Lit2} and linear combinations of  \eqref{Lit} and \eqref{Lit2}.

\section{Conclusion}

This paper has shown, formally and in numerical examples, that price-impact can have material effects on asset pricing via an amplification effect on imperfect risk sharing.  Calibrated price-impact helps resolve both the interest rate and volatility puzzles and has a small effect on the equity premium.  In addition, we conjecture that the introduction of jumps would increase the effect of price-impact on the equity premium.

\appendix 

\section{Auxiliary ODE result} 

In the following ODE existence proof, there are no restrictions on the time horizon $T\in(0,\infty)$ and the constant $C_0\in\R$. We note that the ODE \eqref{g} is quadratic in $g(t)$ and that the square coefficient $-\frac{2}\alpha$ is negative because $\alpha>0$ is the temporary price-impact due to orders $\theta'_{i,t}$.

\begin{proposition} \label{ODE_main}
For $I\in\N$, $C_0\in\R$,  and positive constants $T,a,\sigma_D,\alpha,k>0$ there exists a unique constant $\hh_0 \in(0,k)$ such that the ODE system: 
\begin{align}
&h'(t)=\frac{2 g(t)}{\alpha}h(t),\quad  h(0)=h_0, \label{h}\\
&f'(t)=1+f(t) \left( \tfrac{a^2 \sigma_D^2}{2I}h(t)-C_0 \right),\quad f(0)=1,  \label{f}\\
&g'(t)=a \sigma_D^2 f(t)- \frac{2 }{\alpha}g(t)^2 + g(t) \left( \tfrac{a^2 \sigma_D^2}{2I}h(t)-C_0 \right),\quad g(0)=0, \label{g}
\end{align}
with initial condition $h_0:=\hh_0$, has a unique solution for $t\in[0,T]$ that satisfies $h(T)=k$.

\end{proposition} 

\proof \ \\
\noindent {\bf Step 1/3 ($h$'s range):} Let $h_0\in (0,k)$ be given. We evolve the ODEs \eqref{h}-\eqref{g} from $t=0$ to the right ($t>0$). The local Lipschitz property of the ODEs ensure that there exists a maximal interval of existence $[0, \tau)$ with $\tau\in (0,\infty]$ by the Picard-Lindel\"of theorem (see, e.g., Theorem II.1.1 in Hartman 2002).

For a constant $c$, let $T_{f=c}\in[0,\tau]$ be defined as
\begin{align}
T_{f=c}:=\inf \left\{ t \in (0, \tau): \, f(t)=c  \right\}\land \tau,
\end{align}
where --- as usual --- the infimum over the empty set is defined as $+\infty$. We define $T_{g=c}$ and $T_{h=c}$ similarly. Suppose that $T_{f=0}<\tau$. Then,  $f(0)=1$ and the continuity of $f$ imply that $f(t)>0$ for $t\in [0,T_{f=0})$. Since $f(T_{f=0})=0$, we have $f'(T_{f=0})\leq 0$, but \eqref{f} implies $f'(T_{f=0})=1>0$. Therefore, we conclude that
\begin{align}
T_{f=0}=\tau \quad \textrm{and} \quad f(t)>0  \textrm{  for  }  t\in [0,\tau). \label{f=0}
\end{align}
Because $g(0)=0$ and $g'(0)=a \sigma_D^2>0$, we have $T_{g=0}>0$ and $g(t)>0$ for $t\in (0,T_{g=0})$. The ODE \eqref{h} with $h(0)=h_0>0$ implies that $h(t)$ increases on the interval $[0, T_{g=0})$. Therefore, the ODE \eqref{f} and the positivity of $(f,h)$ produce 
\begin{align}
f'(t)>1-f(t)C_0,\quad t\in [0,T_{g=0}).
\end{align}
Then, Gronwall's inequality produces
\begin{align}  
f(t) \geq 
\begin{cases}  
\frac{1 +(C_0-1)e^{-C_0 t}}{C_0} &\textrm{if  }C_0\neq0, \\   1+t &\textrm{if  }C_0=0.    
\end{cases}
\end{align}
This inequality implies that 
\begin{align}
f(t) \geq C_1 \quad \textrm{for} \quad t\in [0,T_{g=0}) \quad \textrm{where} \quad C_1:=
\begin{cases}
1, &\textrm{if  }C_0\leq 1,\\
\frac{1}{C_0}, &\textrm{if  } C_0>1.
\end{cases} \label{C_1}
\end{align}

Suppose that $T_{g=0}<\tau$. Since $g(t)>0$ for $t\in (0,T_{g=0})$ and $g(T_{g=0})=0$, we have $g'(T_{g=0})\leq 0$. However, this is a contradiction because \eqref{g} and \eqref{C_1} imply $g'(T_{g=0})\geq a \sigma_D^2 C_1 >0$ where the positive constant $C_1$ is defined in \eqref{C_1}.

Up to this point we have shown
\begin{align} \label{g=0}
T_{g=0}=\tau \quad \textrm{and} \quad 
\begin{cases}
f(t)\geq C_1>0, \\ 
h(t)\geq 0, \\ 
g(t) \geq 0, 
\end{cases}  \textrm{  for  }  t\in [0,\tau).
\end{align}
To proceed, the positive constant
\begin{align}
C_2:=\begin{cases}
-\frac{\alpha C_0}{2} ,  & \textrm{if  } C_0<0,\\
\frac{\alpha \left( -C_0+ \sqrt{C_0^2  + \tfrac{4 a \sigma_D^2 C_1}{\alpha} }\right)}{4} , & \textrm{if  }  C_0\geq 0   \label{C_2}
\end{cases}
\end{align}
satisfies
\begin{align}
-\frac2{\alpha} x^2 -C_0 x \geq  - \frac{a \sigma_D^2 C_1}{2}\quad \textrm{for} \quad x\in [0,C_2]. \label{g>C_2} 
\end{align}
Because $0\leq g(t) < C_2$ for $t\in [0, T_{g=C_2})$, we can bound \eqref{g} from below using  \eqref{g=0}  and \eqref{g>C_2} to see for $t\in [0, T_{g=C_2})$
\begin{align}\label{g'}
\begin{split}
g'(t)&\geq a \sigma_D^2C_1- \frac{2 g(t)^2}{\alpha} - g(t) C_0 \\
&\ge \frac12a \sigma_D^2 C_1.
\end{split}
\end{align}
By integrating \eqref{g'} and using the initial condition $g(0)=0$ we see $g(t)\ge \frac12a \sigma_D^2 C_1t$ for $t\in [0, T_{g=C_2})$. Therefore, 
\begin{align}
T_{g=C_2}\leq \tfrac{2C_2}{a \sigma_D^2 C_1}. \label{g=C_2}
\end{align}

Suppose that $T_{g=C_2}=\tau$. Then, for $t\in [0,\tau)$, we have $0 \leq g(t)<C_2$ and the ODE \eqref{h}  produces
\begin{align}\label{hbdd}
\begin{split}
h'(t)&\leq \frac{2 C_2}{\alpha}h(t), \\
h(t) &\leq  h_0 \, e^{\frac{2C_2}{\alpha}t},  
\end{split}
\end{align} 
where the second inequality uses Gronwall's inequality. Similarly, for $t\in [0,\tau)$, the ODE \eqref{f} and Gronwall's inequality imply
\begin{align}\label{fbdd}
\begin{split}
f'(t) &\leq 1+ f(t)\left( \tfrac{a^2 \sigma_D^2}{2I} h(t)+|C_0| \right)\\
 &\leq 1+ f(t)\left( \tfrac{a^2 \sigma_D^2 h_0}{2I}  e^{\frac{2C_2}{\alpha}t}+|C_0| \right),\\
f(t) &\leq (1+t)\exp\left(  |C_0| t + \tfrac{a^2 \sigma_D^2 h_0 \alpha}{4 I C_2}(e^{\frac{2C_2}{\alpha}t}-1)   \right).
\end{split}
\end{align} 
The boundedness properties $g(t)<C_2$, \eqref{hbdd}, and \eqref{fbdd}  imply that $h,f,$ and $g$ do not blow up for $t$ finite. Then, Theorem II.3.1 in Hartman (2002) ensures $\tau=\infty$ which contradicts \eqref{g=C_2}. Consequently, we cannot have $T_{g=C_2}=\tau$ and it must be the case that 
\begin{align}
T_{g=C_2}<\tau. \label{T_g=C_2}
\end{align}

Let $\hT_{g=C_2}$ be defined as the first time $g$  reaches $C_2$ strictly after time $t=T_{g=C_2}$; that is,
\begin{align}
\hT_{g=C_2}:=\inf \big\{ t \in (T_{g=C_2}, \tau): \,\, g(t)=C_2 \big\}\land \tau.
\end{align}
Because $g'(T_{g=C_2})\geq \tfrac{a \sigma_D^2 C_1}{2}>0$ by \eqref{g'}, we have 
\begin{align}
T_{g=C_2}<\hT_{g=C_2} \textrm{  and  }  g(t)> C_2 \textrm{  for  } t\in (T_{g=C_2},\hT_{g=C_2}). \label{hT}
\end{align}
Suppose that $\hT_{g=C_2}<\tau$. Then, $g(\hT_{g=C_2})=C_2$ and \eqref{hT} imply that $g'(\hT_{g=C_2})\leq 0$, but \eqref{g}, \eqref{g=0}, and \eqref{g>C_2} produce the contradiction: 
\begin{align}
\begin{split}
g'(\hT_{g=C_2}) &=a \sigma_D^2 f(\hT_{g=C_2})- \frac{2 C_2^2}{\alpha} +C_2 \left( \tfrac{a^2 \sigma_D^2}{2I}h(\hT_{g=C_2})-C_0 \right)\\
&\geq a \sigma_D^2 C_1- \frac{2 C_2^2}{\alpha} - C_2 C_0  \\
&\geq \tfrac{a \sigma_D^2 C_1}{2}\\
&>0.
\end{split}
\end{align}
Therefore, it must be the case that $\hT_{g=C_2}=\tau$, which implies the lower bound
\begin{align}
g(t)\geq C_2>0  \quad \textrm{for}\quad t\in [T_{g=C_2},\tau). \label{g lower bound}
\end{align}
Combining \eqref{g'} and \eqref{g lower bound} gives the following global lower bound:
\begin{align}\label{g lower bound2}
g(t)\geq  \tfrac{a \sigma_D^2 C_1}{2}t\wedge C_2 \quad \textrm{for} \quad t\in [0,\tau).
\end{align}
In turn, using the ODE \eqref{h}, the bound \eqref{g lower bound2} produces the global lower bound for $h$ via Gronwall's inequality:
\begin{align} \label{h lower bound}
h(t)\geq h_0 \exp\left(\frac{2}{\alpha}\int_0^t  \tfrac{a \sigma_D^2 C_1}{2} s \land C_2  ds 
\right)\quad \textrm{for} \quad t\in [0,\tau).
\end{align}
Next, we suppose $T_{h=k}=\tau$. Then, for $t\in [0,\tau)$, we have $0 \leq h(t)<k$, and the ODEs \eqref{f}-\eqref{g} and Gronwall's inequality imply
\begin{align} \label{f g upper}
\begin{split}
f'(t) &\leq 1+ f(t)C_3,\\
f(t) &\leq (1+t) e^{C_3 t},  \\
g'(t)&\leq a \sigma_D^2 f(t) + g(t) C_3\\
&\leq a \sigma_D^2  (1+t) e^{C_3 t}+ g(t) C_3,\\
g(t)&\leq a \sigma_D^2e^{C_3 t}(t+\tfrac12t^2),
\end{split}
\end{align} 
where $C_3:= \tfrac{a^2 \sigma_D^2}{2I}k +|C_0| $. The inequalities in \eqref{f g upper} and $0 \leq h(t)\le k$ imply that $h,f,$ and $g$ do not blow up for $t$ finite. Then, Theorem II.3.1 in Hartman (2002) ensures $\tau=T_{h=k}=\infty$. This is a contradiction because \eqref{h lower bound} implies that $h(t)$ reaches $k$ in finite time. Therefore, it must be the case that
\begin{align}
T_{h=k}<\tau. \label{T_h finite}
\end{align}

\noindent {\bf Step 2/3 (Monotonicity):} Let $0<h_0<\tilde{h}_0<k$, and denote the solution of the ODE system \eqref{h}-\eqref{g} with the initial condition $h(0)=\tilde{h}_0$ by $\tilde{f},\tilde{h},$ and $\tilde{g}$. The corresponding maximal existence interval is denoted by $\tilde{\tau}$. We define $T_{g=\tilde{g}}$ as
\begin{align}
T_{g=\tilde{g}}:= \inf \left\{ t \in (0, \tau\wedge \tilde{\tau}): \, g(t)=\tilde{g} (t) \right\}\land  \tau\wedge \tilde{\tau}.
\end{align}
Because $g(0)=\tilde{g}(0)=0$, the ODEs \eqref{h}-\eqref{g} have the properties $g'(0)=\tilde{g}'(0)=a \sigma_D^2$ and
\begin{align*}
g''(0)&=a \sigma_D^2 \left( 1+ \tfrac{a^2 \sigma_D^2}{I}h_0 -2 C_0   \right) \\&
<a \sigma_D^2 \left( 1+ \tfrac{a^2 \sigma_D^2}{I}\tilde{h}_0 -2 C_0   \right)\\
&=\tilde{g}''(0).
\end{align*}
Therefore,
\begin{align}\label{g compare}
0<g(t)<\tilde{g}(t) \quad \textrm{for} \quad t\in (0, T_{g=\tilde{g}}).
\end{align}

Suppose that $T_{g=\tilde{g}}<\tau\wedge \tilde{\tau}$. The inequality \eqref{g compare} and the ODEs \eqref{h} and \eqref{f} imply that
\begin{align}
\begin{cases}h(t)<\tilde{h}(t)\\ 
f(t)<\tilde{f}(t)
\end{cases}
 \quad \textrm{for} \quad t\in (0, T_{g=\tilde{g}}]. \label{f h compare}
\end{align}
Also, \eqref{g compare} and $g(T_{g=\tilde{g}})=\tilde{g}(T_{g=\tilde{g}})$ produce $g'(T_{g=\tilde{g}})\geq \tilde{g}'(T_{g=\tilde{g}})$. However, this contradicts 
\begin{align}
\begin{split}
g'(T_{g=\tilde{g}})&= a \sigma_D^2 f(T_{g=\tilde{g}})- \frac{2 g(T_{g=\tilde{g}})^2}{\alpha} + g(T_{g=\tilde{g}}) \left( \tfrac{a^2 \sigma_D^2}{2I}h(T_{g=\tilde{g}})-C_0 \right)\\
&<a \sigma_D^2 \tilde{f}(T_{g=\tilde{g}})- \frac{2 \tilde{g}(T_{g=\tilde{g}})^2}{\alpha} + \tilde{g}(T_{g=\tilde{g}}) \left( \tfrac{a^2 \sigma_D^2}{2I}\tilde{h}(T_{g=\tilde{g}})-C_0 \right)\\&
=\tilde{g}'(T_{g=\tilde{g}}),
\end{split}
\end{align}
where we used \eqref{g} and \eqref{f h compare}. Therefore, we conclude that  $T_{g=\tilde{g}}=\tau \wedge \tilde{\tau}$  and
\begin{align}
\begin{cases}h(t)<\tilde{h}(t)\\ 
f(t)<\tilde{f}(t)\\
g(t)<\tilde{g}(t)
\end{cases}
 \quad \textrm{for} \quad t\in (0,\tau\wedge \tilde{\tau}). \label{f h compare2}
\end{align}

\noindent{\bf Step 3/3 (Existence):} To emphasize the dependence on the initial condition $h(0)=h_0$, we write $\tau(h_0)$ and $T_{h=k}(h_0)$. For example,
\begin{align}
T_{h=k}(h_0):=\inf \left\{ t \in \big(0, \tau(h_0)\big): \, h(t)=k \right\}\land  \tau(h_0).
\end{align}
Inequality \eqref{T_h finite} in Step 1 implies that $T_{h=k}(h_0)<\infty$ for $h_0\in (0,k)$. Step 2 implies that the map $(0,k)\ni h_0\mapsto T_{h=k}(h_0)$ is strictly decreasing. Therefore, the following three statements and the Intermediate Value Theorem complete the proof in the sense that we can choose a unique $\hat{h}_0\in (0,k)$ such that $T_{h=k}(\hat{h}_0)=T$ (recall that $T\in(0,\infty)$ is the model time horizon):
\begin{itemize}
\item[(i)] $\lim_{h_0 \uparrow k}T_{h=k}(h_0)=0$.
\item[(ii)] $\lim_{h_0 \downarrow 0} T_{h=k}(h_0)= \infty$.
\item[(iii)] The map $(0, k)\ni h_0 \mapsto T_{h=k}(h_0)$ is continuous.  
\end{itemize}
Here are the proofs of these three statements:

(i) Inequality \eqref{h lower bound} implies (i).

(ii) The inequalities in \eqref{f g upper} and Gronwall's inequality produce
\begin{align}\label{h upper}
\begin{split}
h(t) &  =  h_0  \exp \left( \int_0^t \tfrac{2 g(s)}{\alpha}  ds    \right) \\
&\leq h_0  \exp \left( \int_0^t \tfrac{2a \sigma_D^2e^{C_3 s}(s+\tfrac12s^2)}{\alpha}  ds    \right).
\end{split}
\end{align} 
Obviously, the function $[0,\infty)\ni t \to\exp \big( \int_0^t \tfrac{2a \sigma_D^2e^{C_3 s}(s+\tfrac12s^2)}{\alpha}  ds\big)$ is increasing. Therefore, for any $t_0>0$, we can choose $h_0>0$ such that
$$
h_0<k\exp\left( -\int_0^t \tfrac{2a \sigma_D^2e^{C_3 s}(s+\tfrac12s^2)}{\alpha}  ds\right),\quad t \in [0,t_0],
$$
and use \eqref{h upper} to see $T_{h=k}(h_0)>t_0$. This shows (ii).

(iii) Let $h_0\in (0,k)$ be fixed. To emphasize the dependence on the initial condition, we write $\big(h(t),g(t)\big)$ as $\big(h(t,h_0),g(t,h_0)\big)$. The local Lipschitz structure of the ODEs \eqref{h}-\eqref{g} gives us the continuous dependence of their solutions on the initial condition $h_0$ (see, e.g., Theorem V.2.1 in Hartman 2002); that is,
\begin{align} \label{initial continuity}
\lim_{x\to h_0} h(t,x) = h(t,h_0),\quad t\in \big[0,\tau(h_0)\big).
\end{align}

For $0<x< h_0$ we have $T_{h=k}(h_0)<T_{h=k}(x)$, and the ODE \eqref{h} and the Fundamental Theorem of Calculus produce:
\begin{align} \label{right of h_0}
\begin{split}
k &= h\big(T_{h=k}(x),x\big)\\
&= h\big(T_{h=k}(h_0),x\big)+ \int_{T_{h=k}(h_0)}^{T_{h=k}(x)} \tfrac{\partial}{\partial t}h(t, x)dt \\
&=  h\big(T_{h=k}(h_0),x\big)+ \frac{2}{\alpha} \int_{T_{h=k}(h_0)}^{T_{h=k}(x)}  g(t,x)h(t,x) dt \\
&\ge  h\big(T_{h=k}(h_0),x\big)+ \frac{2x}{\alpha} \int_{T_{h=k}(h_0)}^{T_{h=k}(x)}  \big( \tfrac{a \sigma_D^2 C_1}{2}t\wedge C_2\big) e^{\frac{2}{\alpha}\int_0^t  \tfrac{a \sigma_D^2 C_1}{2} s \land C_2  ds}dt\\
&\ge  h\big(T_{h=k}(h_0),x\big)+ xC_4 \big(T_{h=k}(x)-T_{h=k}(h_0)\big),
\end{split}
\end{align}
where the second last line uses the bounds \eqref{g lower bound2}  and \eqref{h lower bound} and $C_4>0$ is an irrelevant constant independent of $x$. Letting $x\uparrow h_0$ and using \eqref{initial continuity} produce
\begin{align}\label{limsup}
 \lim_{x\uparrow h_0} T_{h=k}(x)\le T_{h=k}(h_0).
\end{align}
The opposite inequality trivially holds because $T_{h=k}(x)$ is strictly decreasing. Therefore,  \eqref{limsup} holds with equality. Similarly, for $x\in (h_0,\tfrac{k+h_0}{2})$, we have $T_{h=k}(\tfrac{k+h_0}{2})<T_{h=k}(x)<T_{h=k}(h_0)$ and
\begin{align}\label{right of h_0A}
\begin{split}
h\big(T_{h=k}(h_0),x\big)&= h\big(T_{h=k}(x),x\big)+ \int_{T_{h=k}(x)}^{T_{h=k}(h_0)} \tfrac{\partial}{\partial t}h(t, x)dt\\
&=  k+ \int_{T_{h=k}(x)}^{T_{h=k}(h_0)}  g(t,x)h(t,x) dt\\
&\geq  k+x C_5\big(T_{h=k}(h_0)- T_{h=k}(x)\big),
\end{split}
\end{align}
for a constant $C_5$ independent of $x$.  Letting $x\downarrow h_0$ and using \eqref{initial continuity} produce
\begin{align}\label{limsupp}
 \lim_{x\downarrow h_0} T_{h=k}(x)\ge T_{h=k}(h_0).
\end{align}
Again, the opposite inequality trivially holds because $T_{h=k}(x)$ is strictly decreasing. Therefore,  \eqref{limsupp} holds with equality and the continuity property follows.

$\endproof$

\begin{proposition}\label{ODE_sub}
Let $h_0=0$ in \eqref{h}. Then, the ODEs \eqref{h}-\eqref{g} have unique solutions on $t\in [0,\infty)$ with $h(t)=0$ for all $t\geq 0$. 
\end{proposition}
\emph{Proof.} As in the proof of Proposition \ref{ODE_main}, denote the maximal interval of existence by $(0,\tau)$ for $\tau\in (0,\infty]$. For $t\in [0,\tau)$, the solutions to 
 \eqref{h} and \eqref{f} are
 \begin{align} \label{h easy}
 \begin{split}
 h(t)&=0, \\ 
 f(t) &= \begin{cases}  \frac{1+(C_0-1)e^{-C_0 t}}{C_0} &\textrm{if  }C_0\neq0 \\   1+t &\textrm{if  }C_0=0    
 \end{cases}.
 \end{split}
 \end{align}
As in the proof of Proposition \ref{ODE_main}, we can check that 
\begin{align}\label{g easy}
g(t) \geq 0 \quad \textrm{for} \quad t\in [0,\tau). 
\end{align}
Then \eqref{g}, \eqref{h easy}, and \eqref{g easy} imply that for $t\in [0,\tau)$,
\begin{align}
\begin{split}
g'(t) &= a \sigma_D^2 f(t) - \tfrac{2 g(t)^2}{\alpha}- C_0 g(t) \\
&\leq a \sigma_D^2 f(t) + \tfrac{\alpha C_0^2}{8}.
\end{split}
\end{align}
Gronwall's inequality implies that $g$ cannot blow up in finite time. Therefore, we conclude that $\tau=\infty$. 

$\endproof$

\section{Proof of Lemma  \ref{lem:Nash} and Theorem \ref{thm:Nash}}\label{app:pf}

\emph{Proof of  Lemma  \ref{lem:Nash}. } We prove that the coupled ODEs \eqref{ODE_psi}, \eqref{ODE_F}, and \eqref{ODE_Q22} have unique solutions for $t\in[0,T]$. We apply Proposition \ref{ODE_main} and Proposition \ref{ODE_sub} with 
\begin{align}
\begin{split}
C_0 &:=\delta - \frac{a(-2L \mu_D - 2I \mu_Y + 2a L \rho \sigma_D \sigma_Y+a I \sigma_Y^2)}{2I}  - \frac{a^2 \sigma_D^2 L^2}{2I^2},\\
k &:= \sum_{i=1}^I\theta_{i,0}^2-\frac{L^2}I,
\end{split}
\end{align}
where $k$ is non-negative by Cauchy-Schwartz's inequality. The functions
\begin{align}
\begin{split}
\psi(t):=h(T-t)+\tfrac{L^2}{I},\quad F(t):=f(T-t), \quad Q_{22}(t):=-\frac{g(T-t)}{f(T-t)},
\end{split}
\end{align}
solve \eqref{ODE_psi}, \eqref{ODE_F}, and \eqref{ODE_Q22} for $t\in[0,T]$.

From \eqref{C_1} in the proof of Proposition \ref{ODE_main}, we know that $f(t)$ is bounded away from zero for $t\in[0,T]$. Therefore, the solutions to the linear ODEs for $Q(t)$ and $Q_2(t)$ in \eqref{ODE_Q} and \eqref{ODE_Q2} can be found by integration.

$\endproof$
 \ \\

\noindent \emph{Proof of  Theorem  \ref{thm:Nash}. }

\noindent{\bf Step 1/2 (Individual optimality):} In this step, we define the function 
\begin{align}\label{valn}
\begin{split}
&v(t,M_i,D,\theta_i,Y_i):=e^{-a\big(\frac{M_{i}}{F(t)} +D\theta_i+Y_i+ Q(t) + Q_{2}(t)\theta_i+\frac12Q_{22}(t)\theta_i^2\big)},
\end{split}
\end{align}
for $t\in[0,T]$ and $M_i,D,\theta_i,Y_i\in \R$. In  \eqref{valn}, the deterministic functions are defined in \eqref{ODE_F}-\eqref{ODE_Q22}. We note the terminal ODE conditions produce
\begin{align}\label{valnT}
\begin{split}
&v(T,M_i,D,\theta_i,Y_i)=e^{-a(M_{i} +D\theta_i+Y_i)}.
\end{split}
\end{align}
Consequently, because $S_{i,T}=D_T$, we have 
\begin{align}\label{valnTT}
\begin{split}
&e^{-\delta T}v(T,M_{i,T},D_T,\theta_{i,T},Y_{i,T})=e^{-a(X_{i,T}+Y_{i,T})-\delta T},
\end{split}
\end{align}
which is the terminal condition in \eqref{optproblem}. Next, we show that the function $e^{-\delta t} v$ with $v$ defined in \eqref{valn} is the value function for \eqref{optproblem}. To see this, let $(\theta'_i,c_i)\in \sA$ be  arbitrary.  It\^o's lemma shows that the process $e^{-\delta t}v+\int_0^t e^{-a c_{i,u} - \delta u}du$ --- with $v$ being shorthand notation  for the process $v(t,M_{i,t},D_t,\theta_{i,t},Y_{i,t})$ --- has dynamics
\begin{align}\label{dvNash}
\begin{split}
&d\big(e^{-\delta t}v\big)+ e^{-a c_{i,t} - \delta t}dt \\
&=e^{-\delta t}v\Big(
e^{ a (-c_{i,t}+D_t \theta_{i,t} +\frac{M_{i,t}}{F(t)}+Q(t)+\theta_{i,t}  Q_2(t)+\frac{1}{2} \theta_{i,t}^2 Q_{22}(t)+Y)}\\
&-\tfrac{-a \alpha (\theta_{i,t}')^2-a c_{i,t}+a D_t \theta_{i,t} +a Q(t)+a \theta_{i,t}  Q_2(t)+\frac{1}{2} a \theta_{i,t}^2 Q_{22}(t)+a Y-\log \left(\frac{1}{F(t)}\right)+1}{F(t)}
\\
&+\tfrac{a F(t) Q_{22}(t)^2 (L-\theta_{i,t}  I)^2}{\alpha I^2}-\tfrac{a M_{i,t}}{F(t)^2}+\tfrac{2 a \theta_{i,t}' Q_{22}(t) (L-\theta_{i,t}  I)}{I}
 \Big) dt\\
&-ae^{-\delta t}v\Big(\theta_{i,t} \sigma_DdB_t + \sigma_Y\big(\rho dB_t +\sqrt{1-\rho^2}dW_{i,t}\big)\Big),
\end{split}
\end{align}
where we have used the ODEs \eqref{ODE_psi}-\eqref{ODE_Q22} and the interest rate \eqref{r_eq}. The local martingale on the last line in \eqref{dvNash} can be upgraded to a martingale. To see this, we note that $\theta_{i,t}$ is bounded and $v$ is square integrable by \eqref{mmg} so we can use Cauchy-Schwartz's inequality to obtain the needed integrability. 
Furthermore, to see that the drift in \eqref{dvNash} is non-negative, we note the second-order conditions for the HJB equation are (there are no cross terms)
\begin{align}\label{SOC}
\begin{split}
&\theta_{i,t}':\quad \frac{a\alpha}{F(t)}>0,\\
&c_{i,t}:\quad a^2e^{-a c_{i,t}}>0.
\end{split}
\end{align}
This first inequality in \eqref{SOC} holds because $F(t)$ in \eqref{ODE_F} is the annuity ($>0$). Consequently, the drift in \eqref{dvNash} is minimized to zero by the controls \eqref{KLthetahat} and \eqref{KLchat}. This implies that $e^{-\delta t}v+\int_0^t e^{-a c_{i,u} - \delta u}du$ is a submartingale for all admissible order-rate and consumption processes $\theta'_{i,t}$ and $c_{i,t}$. 

It remains to verify admissibility of the controls \eqref{KLthetahat} and \eqref{KLchat}. The explicit solution \eqref{explicitholdings} is deterministic and uniformly bounded. Inserting the controls \eqref{KLthetahat} and \eqref{KLchat} into the money market account balance dynamics \eqref{dM} produces
\begin{align}\label{dM2}
\begin{split}
d M_{i,t} &= \Big( r(t)M_{i,t} +\hat{\theta}_{i,t}D_t -\hat{S}_t\hat{\theta}'_{i,t}+(Y_{i,t} -\hat{c}_{i,t})\Big)dt\\
&=\Big(\tfrac{\log (\tfrac{1}{F(t)})}{a}+M_{i,t} \big(r(t)-\tfrac{1}{F(t)}\big)-Q(t)\\
&-\frac{1}{2} \hat{\theta}_{i,t}  \big(2 Q_{2}(t)+\hat{\theta}_{i,t}  Q_{22}(t)\big)-\hat{S}_{t}\hat{\theta}'_{i,t} \Big)dt.
\end{split}
\end{align}
The linear SDE \eqref{dM2} has a unique well-defined (Gaussian) solution that satisfies  \eqref{mmg}. All in all, this shows the admissibility requirements in Definition \ref{ad} and, hence, optimality of  \eqref{KLthetahat} and \eqref{KLchat} follows from the martingale property of $e^{-\delta t}v+\int_0^t e^{-a \hat{c}_{i,u} - \delta u}du$.\ \\

\noindent{\bf Step 2/2 (Clearing):} Clearly, summing the optimal orders in \eqref{KLthetahat} and using
$\sum_{i=1}^I \theta_{i,0}=L$ show that the stock market clears for all $t\in[0,T]$. Summing \eqref{KLchat} gives us
\begin{align}
\sum_{i=1}^I\hat{c}_{i,t} &=I\tfrac{\log \big(F(t)\big)}{a}+D_tL+IQ(t)+L Q_2(t)+ \frac{1}{2}Q_{22}(t)\sum_{i=1}^I\hat{\theta}_{i,t}^2+\sum_{i=1}^IY_{i,t}.
\end{align}
Because $\psi(0) = \sum_{i=1}^I\theta_{i,0}^2$ and $\sum_{i=1}^I\hat{\theta}_{i,t}^2$ satisfies the ODE \eqref{ODE_psi}, we have $\psi(t)= \sum_{i=1}^I\hat{\theta}_{i,t}^2$ for all $t\in[0,T]$. Therefore, the real good market clears if and and only if
\begin{align}\label{cleeeer1}
0 &=I\tfrac{\log \big(F(t)\big)}{a}+IQ(t)+L Q_2(t)+ \frac12Q_{22}(t)\psi(t).
\end{align}
The terminal conditions in the ODEs \eqref{ODE_Q}-\eqref{ODE_Q22} ensure clearing holds at time $t=T$. By computing time derivatives in \eqref{cleeeer1} and using $r(t)$ defined in \eqref{r_eq}, we see that clearing holds for all $t\in[0,T]$.

Finally, the terminal stock-price condition \eqref{pp10} for the equilibrium stock-price process $\hat{S}_t$ in \eqref{optimal1011} holds by the terminal conditions in the ODEs \eqref{ODE_F}, \eqref{ODE_Q2}, and \eqref{ODE_Q22}.

$\endproof$

\section{Competitive Radner equilibrium}\label{ssecC}

Theorem 2 in Christensen, Larsen, and Munk (2012) shows that there exists a competitive Radner equilibrium in which the equilibrium interest rate is given by
\begin{align}\label{rRad}
\begin{split}
r^\text{Radner} &= \delta +\frac{a}I  (L\mu_D+I\mu_Y)-\frac12\frac{a^2}{I^2}\big(I^2 \sigma_Y^2+2 I L \rho \sigma_D \sigma_Y+L^2 \sigma_D^2\big),
\end{split}
\end{align}
and the equilibrium stock-price process is given by
\begin{align}\label{SRad}
\begin{split}
S^\text{Radner}_t &=\frac{(r^\text{Radner}-1) e^{r^\text{Radner} (t-T)}+1}{r^\text{Radner}}D_t\\
&-\tfrac{\left(e^{r^\text{Radner} (t-T)} ((r^\text{Radner}-1) r^\text{Radner} (t-T)+1)-1\right) \big(\mu_D-\tfrac{a \sigma_D}I (I \rho  \sigma_Y+L \sigma_D)\big)}{(r^\text{Radner})^2}.
\end{split}
\end{align}
It\^o's lemma and \eqref{SRad} produce the competitive Radner equilibrium stock-price volatility coefficient of $S^\text{Radner}_t$ to be
\begin{align}\label{SRadvol}
\begin{split}
\frac{(r^\text{Radner}-1) e^{r^\text{Radner} (t-T)}+1}{r^\text{Radner}}\sigma_D.
\end{split}
\end{align}
Equivalently, we can write \eqref{SRadvol} as $F^\text{Radner}(t)\sigma_D$ where the Radner  annuity $F^\text{Radner}(t)$ is given by \eqref{ODE_FRadner}.

\section{Pareto efficient equilibrium}\label{ssecR}
The following analysis uses the C-CAPM analysis from Breeden (1979). The utilities \eqref{optproblem} produce the representative agent's utility function as
\begin{align}\label{Urep}
-e^{-\frac{a}I c - \delta t},\quad c\in \R,\quad t\in[0,T].
\end{align} 
Because the economy's aggregate consumption is 
$ LD_t + \sum_{i=1}^{I} Y_{i,t}$, the Pareto efficient equilibrium model's unique state-price density $\xi^\text{Pareto}=(\xi_t^\text{Pareto})_{t\in[0,T]}$ is proportional to the process
\begin{align}\label{xiRep}
e^{-\frac{a}I(LD_t+\sum_{i=1}^{I} Y_{i,t}) - \delta t},\quad t\in[0,T].
\end{align}
It\^o's lemma produces the relative state-price dynamics to be:
\begin{align}\label{repagent}
\begin{split}
&\frac{d\xi_t^\text{Pareto}}{\xi_t^\text{Pareto}}\\
&=- \delta dt -\frac{a}I\Big(L dD_t +\sum_{i=1}^{I} dY_{i,t}\Big) +\frac12\frac{a^2}{I^2}d\langle LD+\sum_{i=1}^{I} Y_{i}\rangle_t \\
&= - \delta dt -\frac{a}I \Big((L\mu_D+I\mu_Y) dt + (L\sigma_D+I\sigma_Y\rho)dB_t + \sigma_Y \sqrt{1-\rho^2} \sum_{i=1}^I dW_{i,t}\Big) \\&+\frac12\frac{a^2}{I^2}\Big((L\sigma_D+I\sigma_Y\rho)^2 +I\sigma_Y^2(1-\rho^2)\Big)dt .
\end{split}
\end{align}
From  \eqref{repagent}, the Pareto efficient equilibrium's interest rate (i.e., the $dt$ term in
 $-\frac{d\xi_t^\text{Pareto}}{\xi_t^\text{Pareto}}$) and the market price of risk related to the Brownian motion $B_t$ (i.e., the $dB_t$ volatility term in $-\frac{d\xi_t^\text{Pareto}}{\xi_t^\text{Pareto}}$) are
\begin{align}\label{repagent2}
\begin{split}
r^\text{Pareto} &= \delta +\frac{a}I  (L\mu_D+I\mu_Y)-\frac12\frac{a^2}{I^2}\big((L\sigma_D+I\sigma_Y\rho)^2 +I\sigma_Y^2(1-\rho^2)\big),\\
\lambda&= \frac{a}I (L\sigma_D+I\sigma_Y\rho).
\end{split}
\end{align}
In turn, \eqref{repagent2} produces the  stock-price process in the Pareto efficient equilibrium to be
\begin{align}\label{repagentS}
\begin{split}
S^\text{Pareto}_t &= \frac1{\xi_t^\text{Pareto}}\E_t \Big[\int_t^T D_u\xi_u^\text{Pareto} du + D_T\xi_T^\text{Pareto} \Big]\\
&=-\tfrac{\left(e^{r^\text{Pareto} (t-T)} ((r^\text{Pareto}-1) r^\text{Pareto} (t-T)+1)-1\right) \big(\mu_D-\frac{a \sigma_D}I (I \rho  \sigma_Y+L \sigma_D)\big)}{(r^\text{Pareto})^2}\\&+\tfrac{(r^\text{Pareto}-1) e^{r^\text{Pareto} (t-T)}+1}{r^\text{Pareto}}D_t.
\end{split}
\end{align}
It\^o's lemma and \eqref{repagentS} produce the Pareto efficient equilibrium stock-price volatility coefficient of $S^\text{Pareto}_t$ to be
\begin{align}\label{repagent222}
\frac{(r^\text{Pareto}-1) e^{r^\text{Pareto} (t-T)}+1}{r^\text{Pareto}}\sigma_D.
\end{align}
Equivalently, we can write \eqref{repagent222}
 as $F^\text{Pareto}(t)\sigma_D$ where the annuity $F^\text{Pareto}(t)$ is given by \eqref{ODE_FPareto}.
 
 \section{Transitory Price-Impact Calibration }\label{app:cal}

The challenge in calibrating the transitory price-impact parameter $\alpha$ in \eqref{optimal101B}   is that, $\alpha$ in our model is a measure of the perceived price-impact of fundamental trading imbalances for the aggregate stock market due to frictions in accessing asset-holding capacity from other natural end-counterparties (e.g., large pensions and mutual funds) and not transactional bid-ask bounce and market-maker inventory effects. In contrast, most empirical research measures transitory price effects for individual orders for individual stocks (e.g., as in Hasbrouck (1991) and Hendershott and Menkveld (2014)). The two concepts are related but there are some differences: First, $\alpha$ represents the transitory price effects of sustained trading programs associated with underlying parent orders rather than with isolated child orders (see, e.g., O'Hara (2015)) and one-off single orders. Second, sustained trading occurs in practice both via liquidity-making limit orders as well as via liquidity-taking market orders. From a transactional perspective, market and limit orders have opposite prices of liquidity since one is paying for liquidity and the other is being compensated for providing liquidity. However, limit buying and market buying both create fundamental asset-holding pressure on the available ultimate (i.e., non-market-maker) asset sellers. It is the latter that $\alpha$ measures in our model. Third, stock in our model represents the aggregate stock market as an asset class and, thus, differs from individual stocks both in terms of its scale and as being a source of systematic risk rather than also including idiosyncratic stock-specific randomness.  As a result, it seems  natural, for example, to measure aggregate trading imbalances relative to market capitalization (as a measure of fundamental distortions in aggregate asset supply and demand) rather than in terms of shares (as in a transactional market-maker inventory model).

Our calibration involves adjusting empirical estimates of transitory price-impact for individual stocks into an estimate of the transitory price-impact of trading demand imbalances for the aggregate market. We proceed as follows: First, rather than using price-impact measures for individual trades (e.g., as in Hasbrouck (1991)) or market-maker inventory changes (e.g., as in Hendershott and Menkveld (2014)), we use estimates of the daily transitory price-impact of parent orders in Almgren, Thum, Hauptmann, and Li (2005).  One advantage of the Almgren et al. (2005) estimation for our purposes is that it measures  transitory price-impacts at the parent order level rather than at the child order level.  Another advantage is that there is a natural way to rescale estimated transitory price-impact for individual stocks into a price-impact for the aggregate market.   In particular, the Almgren et al. (2005) estimation is an industry-standard approach in which daily price-impact is estimated given panel data for a sample of parent orders over time for a cross-section of actively traded stocks.  In doing so, the transitory price-impact (TPI) is scaled relative to a stock's individual price and daily return volatility and by scaling the underlying parent order size $\Delta\theta$ as a percentage relative to a stock's average daily trading volume (ADV):

\begin{align}\label{eq:almgrenmodel}
\frac{\text{TPI}}{\text{stock price}\times\text{daily stock return volatility}} =  \eta \times\Big(\frac{\Delta\theta }{\text{ADV}}\times 100\Big)^\beta.
\end{align}
The coefficient $\eta$ is estimated in Almgren et al. (2005) to be 0.141, and the exponent $\beta$ is estimated to be 0.6 (i.e., slightly larger than the standard square-root model). One final advantage is that these estimates are average effects for all stocks rather than being driven by stock-specific differences in the trading environment for a particular stock (e.g., price level, bid-ask spread, institutional vs. retail ownership, market-maker inventory risk due to idiosyncratic stock returns). This gives a ``dimensionless'' standardized measure of transitory price-impact that can then be rescaled for the aggregate market.

Hence, a preliminary estimate of $\alpha$ in our model is:
\begin{footnotesize}
\begin{align}\label{eq:alphacalib}
\begin{split}
\text{TPI} & \approx  \text{market value} \times \text{daily market return volatility} \times \eta \times \frac{Q_3^\beta - Q_1^\beta}{Q_3 - Q_1} \times \frac{\text{SO}}{\text{ADV}}\times \frac{100}{L}\times \frac{\theta'_i}{265} \\
& \approx 3.5 \times 0.2 \sqrt{\frac{1}{265}} \times 0.141\times \frac{1.36^{0.6} - 0.38^{0.6}}{1.36 - 0.38}\times 121.36\times \frac{100}{100} \times \frac{\theta'_i}{265}\\
& \approx  0.0018 \times\theta'_i.
\end{split}
\end{align}
\end{footnotesize}
The following steps were used to derive \eqref{eq:alphacalib}: First, the market value (\$3.50) is set so that the calibrated absolute (dollar) price-impact is roughly consistent with the stock prices our asset-pricing model produces. 
Second, the daily return volatility is set to a ballpark 20\% annual return volatility for the aggregate stock market deannualized for one trading day. 
Third, the power function in \eqref{eq:almgrenmodel} is linearized using its slope between the empirical interquartile values $Q_1$ and $Q_3$ reported in Almgren et al (2005) for the percentage parent-size/ADV ratio.  
Fourth, the ratio $\frac{\Delta \theta}{\text{ADV}}$ is factored for our model as $\frac{\text{SO}}{\text{ADV}} \, \frac{\Delta \theta}{\text{SO}} = 121.36 \frac{\Delta \theta}{L}$ where 121.36 is the empirical average ratio of shares outstanding to ADV for the NYSE and Nasdaq for 2009-2018,\footnote{ From the {\em 2019 SIFMQ Capital Market Fact Book. } }  and where shares outstanding SO $= L=100$ in our model. This rescaling measures parent order size relative to shares outstanding, which, as discussed above, is a natural measure of trade size in our asset-pricing model.
Fifth, the parameter $\alpha$ in \eqref{optimal101B} in our model measures the transitory price-impact relative to the trading rate $\theta_{i,t}'$ (i.e., where $\theta_{i,t}' dt$ is the instantaneous child order flow).  Thus, we write the daily parent order $\Delta \theta$ as 
\begin{equation}
\Delta \theta = \int_0^{\frac{1}{265}} \theta_i' dt = \theta_i' \frac{1}{265}
\end{equation}
in terms of a constant child flow rate $ \theta'_i$ over a trading day (i.e., $\frac{1}{265}$ of a year).

\end{document}